\documentclass[
reprint,
 amsmath,amssymb,
pra,
]{revtex4-1}
\usepackage{dcolumn}
\usepackage{bm}
\usepackage{amsfonts}
\usepackage{amsmath}
\usepackage{amssymb}
\usepackage{float}
\usepackage{graphicx}
\usepackage{subfigure}
\usepackage{epstopdf}
\usepackage{hyperref}

\begin{document}

\title{Reservoir-engineered entanglement in a hybrid modulated three-mode optomechanical system}
\author{Chang-Geng Liao$^{1,2,3}$}
\author{Rong-Xin Chen$^{4}$}
\email[]{chenrxas@tamu.edu}

\author{Hong Xie$^{5}$}
\author{Xiu-Min Lin$^{1,2}$}
\email[]{xmlin@fjnu.edu.cn}

\affiliation{$^{1}$ Fujian Provincial Key Laboratory of Quantum Manipulation and New Energy Materials, College of Physics and Energy, Fujian Normal University, Fuzhou 350117, China}
\affiliation{$^{2}$ Fujian Provincial Collaborative Innovation Center for Optoelectronic Semiconductors and Efficient Devices, Xiamen 361005, China}
\affiliation{$^{3}$ Department of Electronic Engineering, Fujian Polytechnic of Information Technology, Fuzhou, 350003, China}
\affiliation{$^{4}$ Institute for Quantum Science and Engineering and Department of Physics and Astronomy, Texas A\&M University, College Station, TX77843-4242, USA}
\affiliation{$^{5}$ College of JinShan, Fujian Agriculture and Forestry University, Fuzhou 350002, China}

\begin{abstract}
  We propose an effective approach for generating highly pure and strong cavity-mechanical entanglement (or optical-microwave entanglement) in a hybrid modulated three-mode optomechanical system. By applying two-tone driving to the cavity and modulating the coupling strength between two mechanical oscillators (or between a mechanical oscillator and a transmission line resonator), we obtain an effective Hamiltonian where an intermediate mechanical mode acting as an engineered reservoir cools the Bogoliubov modes of two target system modes via beam-splitter-like interactions. In this way, the two target modes are driven to  two-mode squeezed states in the stationary limit. In particular, we discuss the effects of cavity-driving detuning on the entanglement and the purity. It is found that the cavity-driving detuning plays a critical role in the goal of acquiring highly pure and strongly entangled steady states.
\end{abstract}

\flushbottom
\maketitle

\thispagestyle{empty}

\section{introduction}

Theoretical explorations of a quantum optomechanical system began in the 1990s, including several aspects such as the squeezing of light \cite{fabre1994quantum-noise,mancini1994quantum}, quantum non-demolition detection of the light intensity \cite{Jacobs1994Quantum,Pinard1995Quantum}, preparation of nonclassical states \cite{Bose1997Preparation,Mancini1996Ponderomotive,Schwab2005Putting}, and so on. Ever since the optical feed-back cooling scheme based on the radiation-pressure force was first experimentally demonstrated in 1999 \cite{Cohadon1999Cooling}, cavity optomechanics has attracted much interest, and fruitful progress has been made. Apart from its potential applications in building highly sensitive sensors and in testing macroscopic quantum mechanics \cite{Chen2013Macroscopic}, cavity optomechanics can also serve as a light-matter interface to convert information among different systems such as atoms or atomic ensembles \cite{Hammerer2009Strong,Polzik2008Quantum}, Bose-Einstein condensates \cite{Chen2010Classical,Jing2011Quantum}, superconducting solid state qubits \cite{O2010Quantum}, etc.

To date, a variety of experimental optomechanical setups have been reported, e.g., whispering gallery microdisks \cite{Jiang2009High,wiederhecker2009controlling} and microspheres \cite{ma2007radiation-pressure-driven,park2009resolved-sideband}, membranes \cite{thompson2007strong} or nanorods \cite{favero2009fluctuating} inside Fabry-Perot cavities, nanomechanical beam inside a superconducting transmission line microwave cavity \cite{regal2008measuring}, etc. Notably, the hybrid optomechanical system consisting of different physical components possesses the distinct advantages of each component, which may be beneficial for  quantum- information processing (QIP). As experimentally demonstrated by Lee \cite{lee2009cooling} and Winger \cite{winger2011a}, one can manipulate a mechanical nanoresonator via both the opto- and electro-mechanical interactions, which may provide a platform to entangle microwave and optical fields \cite{barzanjeh2011entangling}.

In this paper, we propose an effective approach for generating strong steady-state opto-mechanical entanglement (or optical-microwave entanglement), which is of great importance for both fundamental physics and applications in QIP. For a simple optomechanical system consisting of a laser-driven optical cavity and a vibrating end mirror, the entanglement between the cavity field and the mechanical resonator can be induced by the radiation pressure. However, the amount of created entanglement is largely limited due to environmental noises and the stability constraints of systems \cite{schmidt2012optomechanical}. To enhance the entanglement strength, a feasible way is to apply a suitable time modulation to the driving laser \cite{Mari2009Gently,mari2012opto}. The method is also effective in three-mode \cite{abdi2015entangling,li2015generations,wang2016macroscopic} or four-mode \cite{chen2014enhancement,sr2017} optomechanical systems. Another promising approach for creating strong entanglement or  squeezing is to induce an effective engineered reservoir by pumping the optomechanical systems with  proper blue and red detuned lasers \cite{chen2014enhancement,sr2017,wang2013reservoir,wang2015bipartite,tan2013dissipation,woolley2014two,yang2015generation,Li2015Generation,Kenan,chen2015dissipation}, which is highly attractive from an experimental point of view. As far as we know, previous studies mostly focused on enhancing entanglement between two cavity fields \cite{wang2013reservoir,wang2015bipartite} or two mechanical oscillators \cite{tan2013dissipation,woolley2014two,yang2015generation,Li2015Generation,wang2016macroscopic,chen2014enhancement,sr2017}.
Here, inspired by the approach in Ref.~\cite{woolley2014two}, which has been experimentally demonstrated recently~\cite{arxiv}, we propose to use both time modulation and reservoir engineering techniques to generate highly pure opto-mechanical or optical-microwave entanglement that goes far beyond the entanglement limit based on coherent parametric coupling (i.e., ln2) \cite{vitali2007optomechanical,paternostro2007creating,Mari2009Gently}. In our hybrid three-mode optomechanical system, the intermediate mechanical mode acting as a cooling reservoir and the sum mode of the Bogoliubov modes of the other two system modes are coupled via  the beam-splitter-like interaction. The sum mode in turn is coupled to the difference mode of the Bogoliubov modes. The swap interactions allow both the sum and the difference modes to be cooled via the dissipative dynamics of the intermediate mechanical mode, which is quite different from  Refs.~\cite{wang2013reservoir,wang2015bipartite}. In Refs.~\cite{wang2013reservoir,wang2015bipartite}, only one of the two Bogoliubov modes of the target modes is cooled while the other Bogoliubov mode is a dark mode that is not coupled to the engineered bath and thus can not be cooled. Accordingly, the obtained steady states are two-modes squeezed thermal states. i.e., mixed states. On the contrary, our proposal allows the engineered bath to cool both Bogoliubov modes simultaneously. In this way, we are able to obtain a highly pure and strongly entangled steady state that is vital in the standard continuous-variable teleportation protocol~\cite{braunstein1998teleportation,adesso2005equivalence}. Moreover, unlike the proposal in Ref.~\cite{woolley2014two}, which mainly focuses on the generation of steady-state mechanical-mechanical entanglement in the adiabatic limit, we show that steady opto-mechanical entanglement (or optical-microwave entanglement) can be maximized by choosing the proper ratio of the effective optomechanical couplings. We also discuss the critical role of the effective Bogoliubov-mode coupling (i.e., the frequency detuning between the cavity and the pumping) on the steady-state entanglement and purity, which is not considered in Ref.~\cite{woolley2014two}.

\section{The model}
\begin{figure}[t]
  \centering
  {\includegraphics[width=0.42\textwidth]{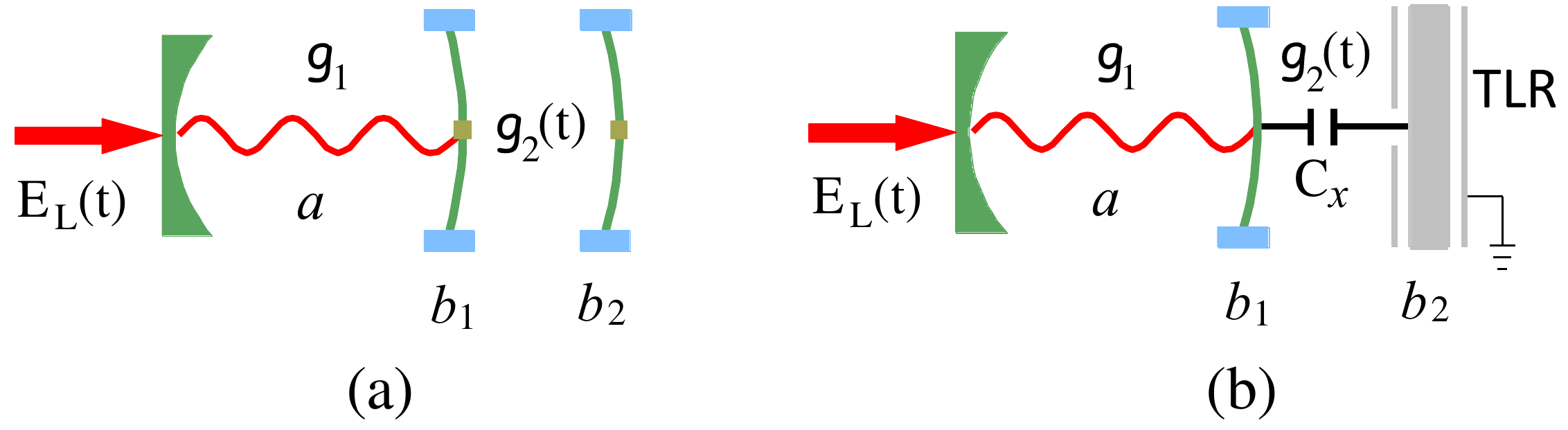}}
  \caption{\label{fig:model}
  (Color online) Schematic representation of the system. An optical cavity mode $a$ driven by a two-tone laser $E_L(t)$ is coupled to an intermediate mechanical mode $b_1$ with single-photon optomechanical coupling strength $g_1$. $b_1$ is in turn coupled, with a time-dependent coupling strength $g_2(t)$, to (a) another mechanical oscillator, or alternatively to (b) a transmission line resonator $b_2$.}
\end{figure}

As shown in Fig.~\ref{fig:model}, a hybrid modulated three-mode optomechanical system is composed of an optical cavity mode $a$ and two mechanical oscillators $b_1$ and $b_2$ [see Fig.~\ref{fig:model}(a)]; or a cavity mode $a$, a mechanical oscillator $b_1$, and a transmission line resonator $b_2$ [see Fig.~\ref{fig:model}(b)]. $g_1$ is   the single-photon optomechanical coupling strength between the cavity mode $a$ with frequency $w_c$ and the intermediate mechanical mode $b_1$ with frequency $w_1$. The cavity is driven by a two-tone laser $E_L(t)$. $g_2(t)$ is the time-dependent coupling between the  intermediate mechanical mode $b_1$  and the second mechanical resonator (or the transmission line resonator) $b_2$ with frequency $w_2$. Here, the controllable mechanical-mechanical coupling $g_2(t)$ in Fig.~\ref{fig:model}(a)
can be realized by using piezoelectrically induced parametric mode mixing \cite{okamoto2013coherent} or by modulating the Coulomb interactions between the mechanical oscillators~\cite{buks2002electrically,Hensinger2005Ion,Zhang2012Precision,Ma2014Tunable,chen2015dissipation}, while  the mechanical-microwave coupling $g_2(t)$ in Fig.~\ref{fig:model}(b) may be achieved via the mechanical displacement-dependent capacitance $C_x$ of the microwave cavity.

The system Hamiltonian reads (set $\hbar=1$)
\begin{eqnarray}
H&=&w_c a^\dag a +w_1b_1^\dag b_1+w_2b_2^\dag b_2+g_1 (b_1+b_1^\dag)a^\dag a\nonumber\\
&&+g_2(t)(b_1+b_1^\dag)(b_2+b_2^\dag)+H_{dr},
\end{eqnarray}
where
\begin{eqnarray}
g_2(t)=&2[g_2^A\cos(w_1+w_2+w_c-w_d)t\nonumber\\
&+g_2^B\cos(w_1-w_2-w_c+w_d)t],
\end{eqnarray}
and $H_{dr}$ is the Hamiltonian of the two-tone driving with  frequencies $w_d\pm w_1$,
\begin{eqnarray}
H_{dr}&=&(\epsilon_+^*e^{iw_1t}+\epsilon_-^*e^{-iw_1t})e^{iw_dt}a+h.c..
\end{eqnarray}
Moving into a rotating frame by performing the unitary transformation $U=\exp\{-i[w_da^\dag a+w_1b_1^\dag b_1+(w_2+w_c-w_d)b_2^\dag b_2]t\}$, we obtain
\begin{eqnarray}\label{HR}
H_R&=&U^\dag HU-iU^\dag\partial U/\partial t\nonumber\\
&=&\delta (a^\dag a -b_2^\dag b_2)+g_1(b_1e^{-iw_1t}+b_1^\dag e^{iw_1t})a^\dag a\nonumber\\
&&+g_2(t)(b_1e^{-iw_1t}+b_1^\dag e^{iw_1t})[b_2e^{-i(w_2+\delta)t}\nonumber\\
&&+b_2^\dag e^{i(w_2+\delta)t}]+[(\epsilon_+^*e^{iw_1t}+\epsilon_-^*e^{-iw_1t})a\nonumber\\
&&+h.c.],
\end{eqnarray}
where $\delta=w_c-w_d$ is the cavity-driving  frequency detuning.
Applying the displacement transformation $a=\bar{a}_+e^{-iw_1t}+\bar{a}_-e^{iw_1t}+d$ to Eq.~(\ref{HR}) in the  strong driving case, we obtain the linearized Hamiltonian by discarding all nonlinear terms of the quantum fluctuations  provided that the single-photon optomechanical coupling $g_1$ is small,
\begin{eqnarray}\label{Hlin}
H_{lin}=H_0+H_1+H_2,
\end{eqnarray}
with
\begin{subequations}
\begin{align}
H_0=&\delta (d^\dag d -b_2^\dag b_2),\\
H_1=&g_1[(\bar{a}_+b_1d+\bar{a}_-b_1d^\dag)+(\bar{a}_+b_1d^\dag\nonumber\\
&+\bar{a}_-b_1d)e^{-2iw_1t}]+h.c.,\\
H_2=&g_2^A \Big{\{}b_1b_2[1+e^{-2i(w_1+w_2+\delta)t}]\nonumber\\
&+b_1b_2^\dag[e^{2i(w_2+\delta)t}+e^{-2iw_1t}] \Big{\}}\nonumber\\
&+g_2^B \Big{\{}b_1b_2[e^{-2i(w_2+\delta)t}+e^{-2iw_1t}]\nonumber\\
&+b_1b_2^\dag[1+e^{-2i(w_1-w_2-\delta)t}] \Big{\}}+h.c.,
\end{align}
\end{subequations}
where  the classical cavity field amplitudes $\bar{a}_\pm$ are assumed to be real
\begin{eqnarray}
\bar{a}_\pm=i\epsilon_{\pm}/(-\kappa/2-i\delta \pm iw_1),
\end{eqnarray}
and $\kappa$ is the cavity decay rate.
If we set $g_1\bar{a}_+=g_2^A=G_+$, $g_1\bar{a}_-=g_2^B=G_-$, under the conditions $w_1, w_2, |w_1-w_2-\delta|\gg G_\pm$, all the non-resonant terms in the linearized Hamiltonian $H_{lin}$ can be effectively neglected under the rotating-wave approximation
\begin{eqnarray}\label{HRWA}
H_{RWA}=\delta(\beta_1^\dag \beta_1-\beta_2^\dag\beta_2)+[G(\beta_1^\dag+\beta_2^\dag)b_1+h.c.],
\end{eqnarray}
where the  Bogoliubove modes $\beta_{1}$ and $\beta_{2}$ are unitary transformations of $d$ and $b_2$,  respectively
\begin{subequations}\label{beta12}
\begin{align}
\beta_1=&s(r)d s^\dag(r)=d\cosh r+b_2^\dag \sinh r,\\
\beta_2=&s(r)b_2 s^\dag (r)=b_2\cosh r+d^\dag \sinh r.
\end{align}
\end{subequations}
Here, $G=\sqrt{G_-^2-G_+^2}$ (we have assumed $G_+ < G_-$ to ensure stability) and $s(r)=\exp[r(db_2-d^\dag b_2^\dag)]$ is the two-mode squeezing operator with the squeezing parameter $r=\tanh^{-1}(G_+/G_-)$. It's clear from Eq.~(\ref{beta12}) that the joint ground state of $\beta_1$ and $\beta_2$ is the two-mode squeezed vacuum state of the cavity mode $d$ and the mechanical mode $b_2$. Introducing the sum mode and the difference mode of Bogoliubov modes
\begin{eqnarray}
\beta_{sum}=(\beta_1+\beta_2)/\sqrt{2},\ \beta_{diff}=(\beta_1-\beta_2)/\sqrt{2},
\end{eqnarray}
then the Hamiltonian in Eq.~(\ref{HRWA}) becomes
\begin{eqnarray}\label{HRWA1}
H_{RWA}=\delta\beta_{sum}^\dag \beta_{diff}+\sqrt{2}G\beta_{sum}^\dag b_1+h.c.,
\end{eqnarray}
which is similar to that of Ref.~\cite{woolley2014two}. Obviously, the sum mode $\beta_{sum}$ is coupled to both the intermediate mechanical mode $b_1$ and the difference mode $\beta_{diff}$ each via a beam-splitter-like interaction. Through the intermediate mechanical mode $b_1$ acting as an engineered reservoir, both the sum and difference modes, i.e., the two Bogoliubove modes $\beta_1$ and $\beta_2$, can be cooled to near ground state, generating two-mode squeezing between the cavity mode $d$ and the mechanical mode $b_2$.

\section{entanglement and purity}

The quantum Langevin equations governing the dynamics of the linearized system can be written as
\begin{subequations}\label{QL}
\begin{align}
\dot d=&i[H_{lin},d]-\frac{\kappa}{2}d+\sqrt{\kappa}d_{in},\\
\dot b_j=&i[H_{lin},b_j]-\frac{\gamma_j}{2}b_j+\sqrt{\gamma_j}b_{j,in},
\end{align}
\end{subequations}
where $\gamma_j$ ($j=1,2$) is the damping rate for the $j$th mechanical oscillator, and $d_{in}$ and $b_{j,in}$ are independent zero mean vacuum input noise operators obeying the following correlation functions
\begin{subequations}\label{noise}
\begin{align}
<d_{in}(t)d_{in}^\dag(t')>=&(\bar{n}_d+1)\delta(t-t'),\\
<d_{in}^\dag(t)d_{in}(t')>=&\bar{n}_d\delta(t-t'),\\
<b_{j,in}(t)b_{j,in}^\dag(t')>=&(\bar{n}_j+1)\delta(t-t'),\\
<b_{j,in}^\dag(t)b_{j,in}(t')>=&\bar{n}_j\delta(t-t')
\end{align}
\end{subequations}
with $\bar{n}_d$ and $\bar{n}_j$ being equilibrium mean thermal occupancies of the cavity and the $j$th mechanical baths, respectively.

Introducing the position and momentum quadratures for the bosonic modes and their input noises
\begin{eqnarray}
Q_o=(o+o^\dag)/\sqrt{2},\ P_o=(o-o^\dag)/(i\sqrt{2}),
\end{eqnarray}
with $o\in\{d,b_1,b_2,d_{in},b_{1,in},b_{2,in}\}$ and the vectors of all quadratures
\begin{subequations}\label{vct}
\begin{align}
R=&[Q_d,P_d,Q_{b_1},P_{b_1},Q_{b_2},P_{b_2}]^T,\\
N=&[\sqrt{\kappa}Q_{d_{in}},\sqrt{\kappa}P_{d_{in}},\sqrt{\gamma_1}Q_{b_{1,in}},\nonumber\\
&\sqrt{\gamma_1}P_{b_{1,in}},\sqrt{\gamma_2}Q_{b_{2,in}},\sqrt{\gamma_2}P_{b_{2,in}}]^T,
\end{align}
\end{subequations}
the linearized quantum Langevin equations~(\ref{QL}) can be written in a compact form
\begin{eqnarray}\label{R}
\dot R=M(t)R+N.
\end{eqnarray}
Here, $M(t)$ is a $6\times6$ time-dependent matrix
\begin{widetext}
\begin{eqnarray}\label{M}
M(t)= \left( {\begin{array}{*{20}{c}}
-\kappa/2&\delta&Im(G_1+G_2)&Re(G_2-G_1)&0&0\\
-\delta&-\kappa/2&-Re(G_2+G_1)&Im(G_2-G_1)&0&0\\
Im(G_1-G_2)&Re(G_2-G_1)&-\gamma_1/2&0&Im(G_3+G_4)&Re(G_4-G_3)\\
-Re(G_2+G_1)&-Im(G_1+G_2)&0&-\gamma_1/2&-Re(G_3+G_4)&Im(G_4-G_3)\\
0&0&Im(G_3-G_4)&Re(G_4-G_3)&-\gamma_2/2&-\delta\\
0&0&-Re(G_3+G_4)&-Im(G_3+G_4)&\delta&-\gamma_2/2
\end{array}}\right),
\end{eqnarray}
\end{widetext}
where $Re$ and $Im$ respectively denote the real and imaginary parts. $G_1\sim G_4$ are given by
{\allowdisplaybreaks
\begin{subequations}
\begin{align}
G_1=&G_++G_-e^{2iw_1t},\\
G_2=&G_-+G_+e^{-2iw_1t},\\
G_3=&G_+[1+e^{2i(w_1+w_2+\delta)t}]\nonumber\\
&+G_-[e^{2i(w_2+\delta)t}+e^{2iw_1t}],\\
G_4=&G_-[1+e^{2i(w_1-w_2-\delta)t}]\nonumber\\
&+G_+[e^{-2i(w_2+\delta)t}+e^{2iw_1t}].
\end{align}
\end{subequations}}

Since the system is linearized, it remains Gaussian starting from an initial Gaussian state whose information-related properties can be fully described by the covariance matrix \cite{adesso2007entanglement,weedbrook2012gaussian,olivares2012quantum}. For our three-mode bosonic system, the covariance matrix $\sigma$ is a  $6\times 6$ matrix with components defined as
\begin{eqnarray}\label{sigma}
\sigma_{j,k}=<R_jR_k+R_kR_j>/2,
\end{eqnarray}
where $R_k$ is the $k$th component of the vector of  quadratures $R$ in Eq.~(\ref{vct}).
From Eqs.~(\ref{noise}), (\ref{vct}), and (\ref{R}), we can derive a linear differential equation of the covariance matrix that is equivalent to the quantum Langevin equation~(\ref{R}) when only Gaussian states are relevant~\cite{Mari2009Gently},
\begin{eqnarray}\label{dotsigma}
\dot \sigma=M(t)\sigma+\sigma M(t)^T+D.
\end{eqnarray}
Here, $D$ is a diffusion matrix whose components are associated with the noise correlation functions (see Eq.~(\ref{noise}))
\begin{flalign}\label{D}
D_{j,k}\delta(t-t')=<N_j(t)N_k(t')+N_k(t')N_j(t)>/2.
\end{flalign}
$D$  is found to be diagonal,
 \begin{eqnarray}\label{D}
D &=& {\rm{diag}}\{\kappa(2\bar{n}_d+1)/2,\kappa(2\bar{n}_d+1)/2,\gamma_1(2\bar{n}_1+1)/2,\nonumber\\
&& \gamma_1(2\bar{n}_1+1)/2, \gamma_2(2\bar{n}_2+1)/2, \gamma_2(2\bar{n}_2+1)/2\}.
\end{eqnarray}

The general stability conditions of the linear differential equation (Eq.~(\ref{R}) or equally Eq.~(\ref{dotsigma})) are determined by the corresponding homogeneous equation $\dot{R}=M(t)R$, which is fully characterized by the time-periodic coefficient matrix $M(t)$. Suppose that the period of the coefficient matrix $M(t)$  is $T>0$, i.e. $M(t)=M(t+T)$. Let $\Pi(t)$ be a principal matrix solution of the homogeneous equation. The eigenvalues $\lambda_{j}$ ($j=1,2,...,6$) of $\Lambda=\Pi^{-1}(0)\Pi(T)$ are called the characteristic multipliers or Floquet multipliers~\cite{Teschl2012}, where $\Pi(T)$ can be obtained by numerical integration with the initial condition $\Pi(0)$. The solutions of Eq.~(\ref{R}) and Eq.~(\ref{dotsigma}) are stable  if all Floquet multipliers satisfy $|\lambda_{j}|<1$.  For the special case of a time-independent coefficient matrix $M=M(t=0)$ under the  rotating-wave approximation, i.e. omitting  all nonresonant terms in Eq.~(\ref{Hlin}) (all time-dependent terms in Eq.~(\ref{M})), the stability requirements can be readily inferred from the
eigenvalues of the time-independent coefficient matrix $M$, i.e. all eigenvalues of $M$ having negative real parts. The stability conditions will be carefully checked in all simulations throughout this paper.

For two-mode Gaussian states of the cavity mode $d$ and the mechanical resonator $b_2$ of interest here, it is convenient to use the logarithmic negativity $E_N$ as a measurement of the entanglement \cite{plenio2005logarithmic,vidal2002computable}.
$E_N$ can be computed from the reduced $4\times 4$  covariance matrix $\sigma_r$ for $d$ and $b_2$ whose components are just the terms associated with $d$ and $b_2$ only in the full covariance matrix $\sigma$. If we write $\sigma_r$ in the form
\begin{eqnarray}\label{Rs}
\sigma_r=\left( {\begin{array}{*{20}{c}}
V_1&V_c\\
V_c^T&V_2
\end{array}}\right),
\end{eqnarray}
where $V_1,V_2$, and $V_c$ are $2\times 2$ subblock matrices of $\sigma_r$, the logarithmic negativity $E_N$ is then given by
\begin{eqnarray}\label{EN}
E_N=\max[0,-\ln(2\eta)],
\end{eqnarray}
with
\begin{subequations}
\begin{align}
\eta=&2^{-1/2}\{\Sigma-[\Sigma^2-4\det\sigma_r]^{1/2}\}^{1/2},\\
\Sigma=&\det V_1+\det V_2-2\det V_c.
\end{align}
\end{subequations}
The purity of a two-mode Gaussian state described by a covariance matrix $\sigma_r$ is simply given by
\begin{eqnarray}\label{EN}
\mu =1/(4\sqrt{\det \sigma_r}).
\end{eqnarray}

\begin{figure}[t]
   \centering
   \subfigure{\includegraphics[width=0.42\textwidth]{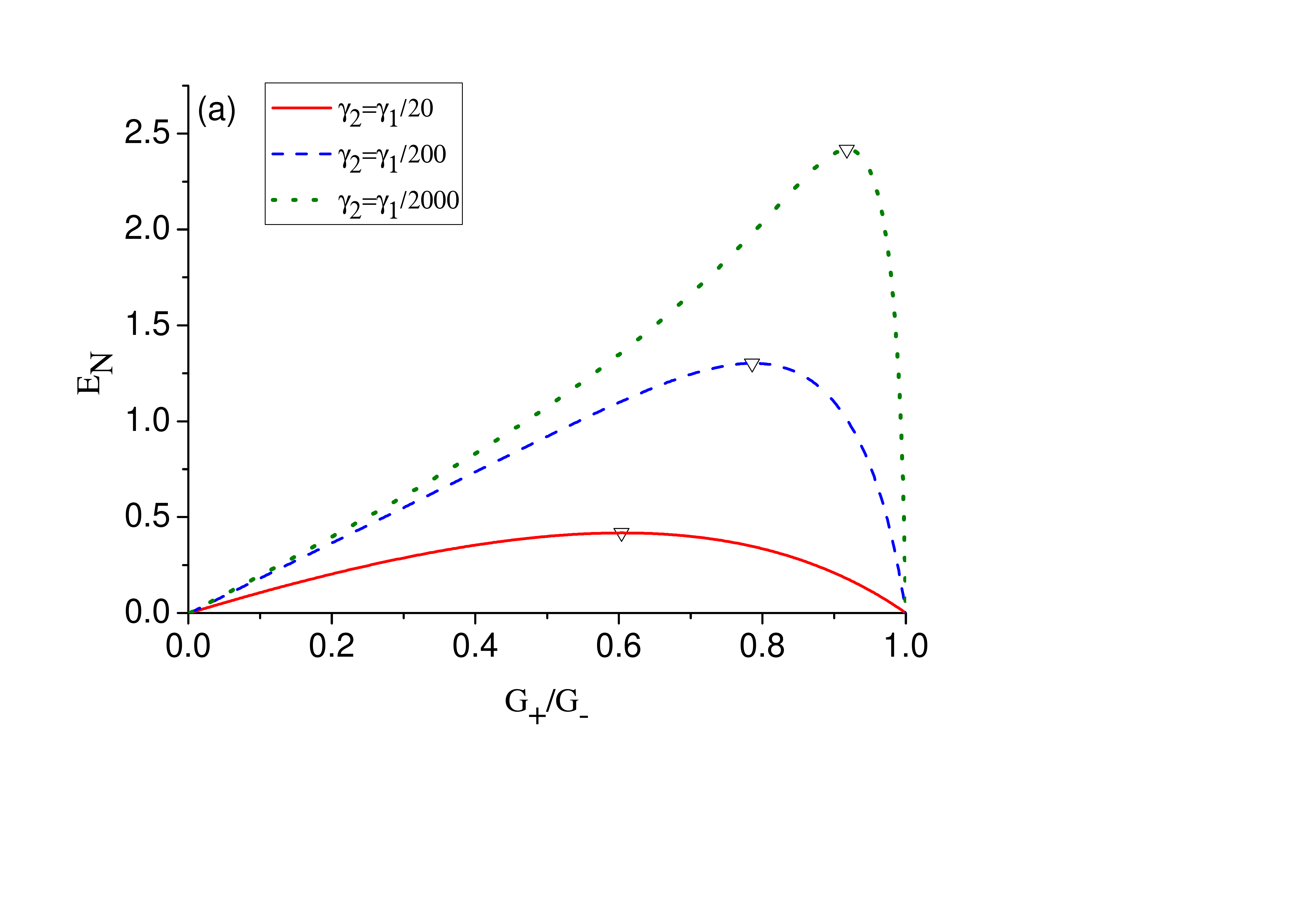}\label{fig:EVCA}}
   \subfigure{\includegraphics[width=0.42\textwidth]{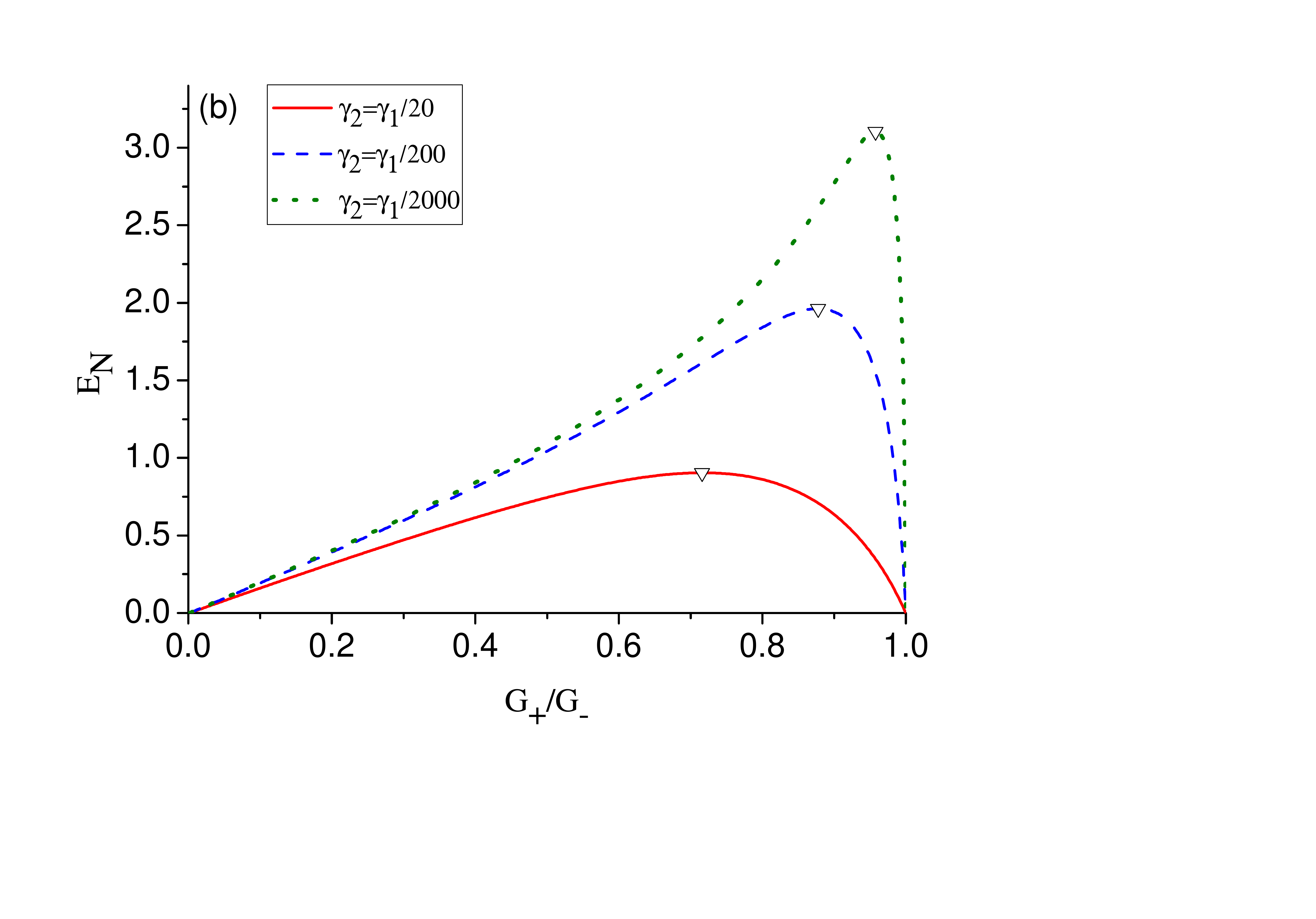}\label{fig:EVCB}}
   \subfigure{\includegraphics[width=0.42\textwidth]{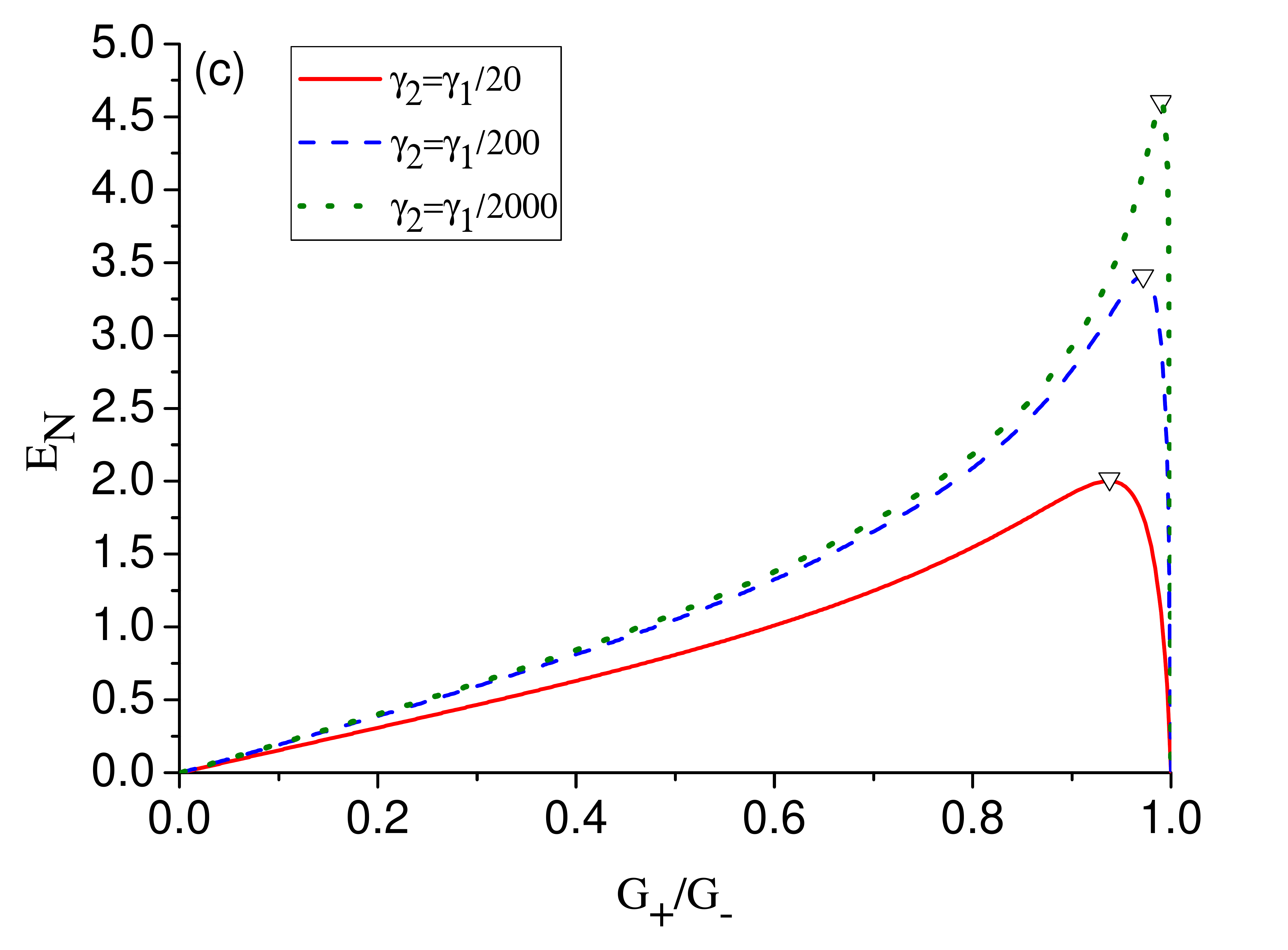}\label{fig:EVCC}}
   \caption{\label{fig:EVC}
   (Color online) Stationary cavity-mechanical entanglement $E_N$ versus the ratio of the effective couplings $G_+/G_-$ for different values of $\delta$. (a) $\delta =10\gamma_1$, (b) $\delta =5\gamma_1$, and (c) $\delta =\gamma_1$. The other parameters are: $G_-=2.5\gamma_1$, $\kappa=\gamma_2$, $\bar{n}_1=\bar{n}_2=\bar{n}_d=0$.}
\end{figure}

\begin{figure}[t]
   \centering
   \subfigure{\includegraphics[width=0.42\textwidth]{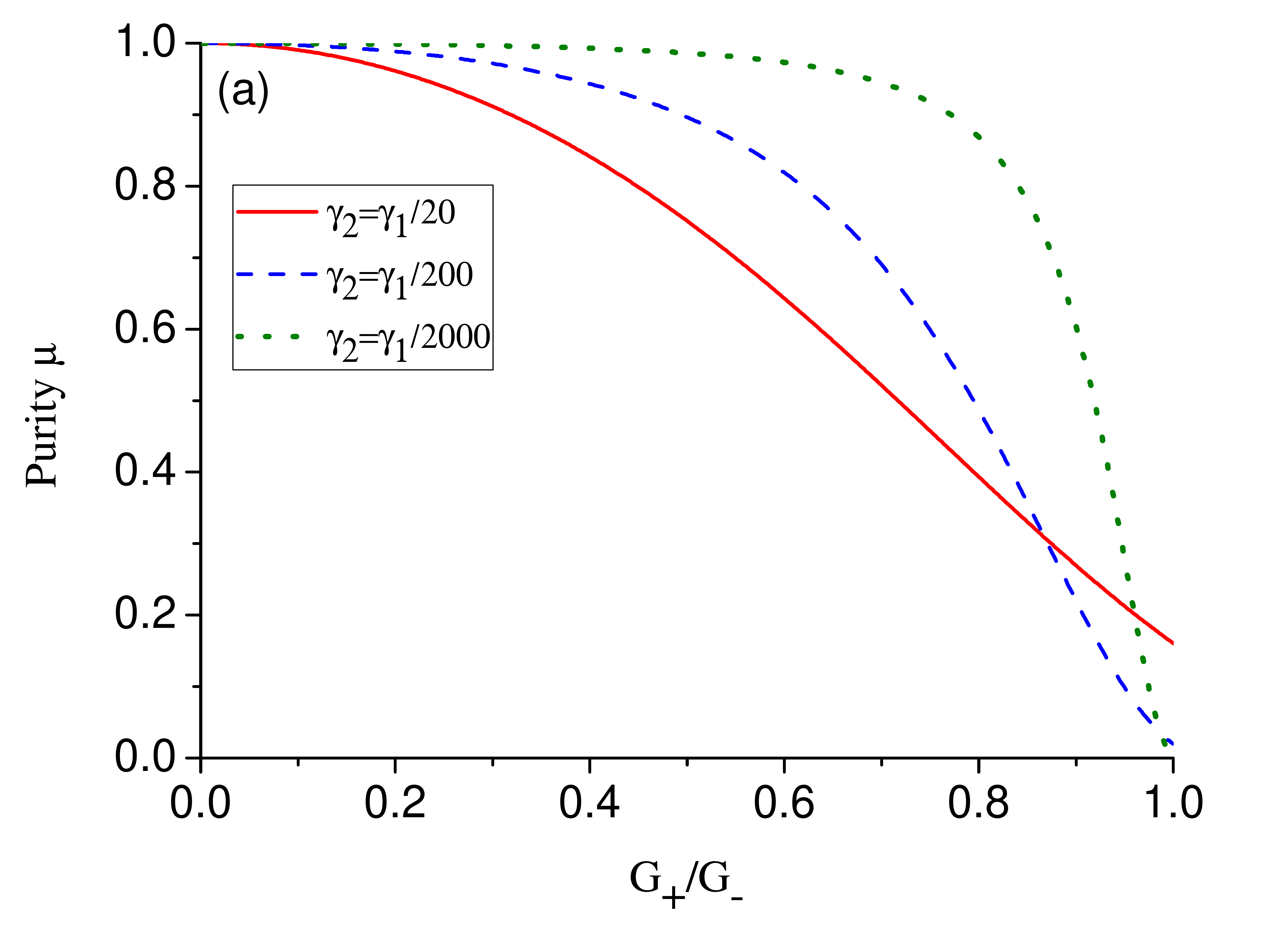}\label{fig:PA}}
   \subfigure{\includegraphics[width=0.42\textwidth]{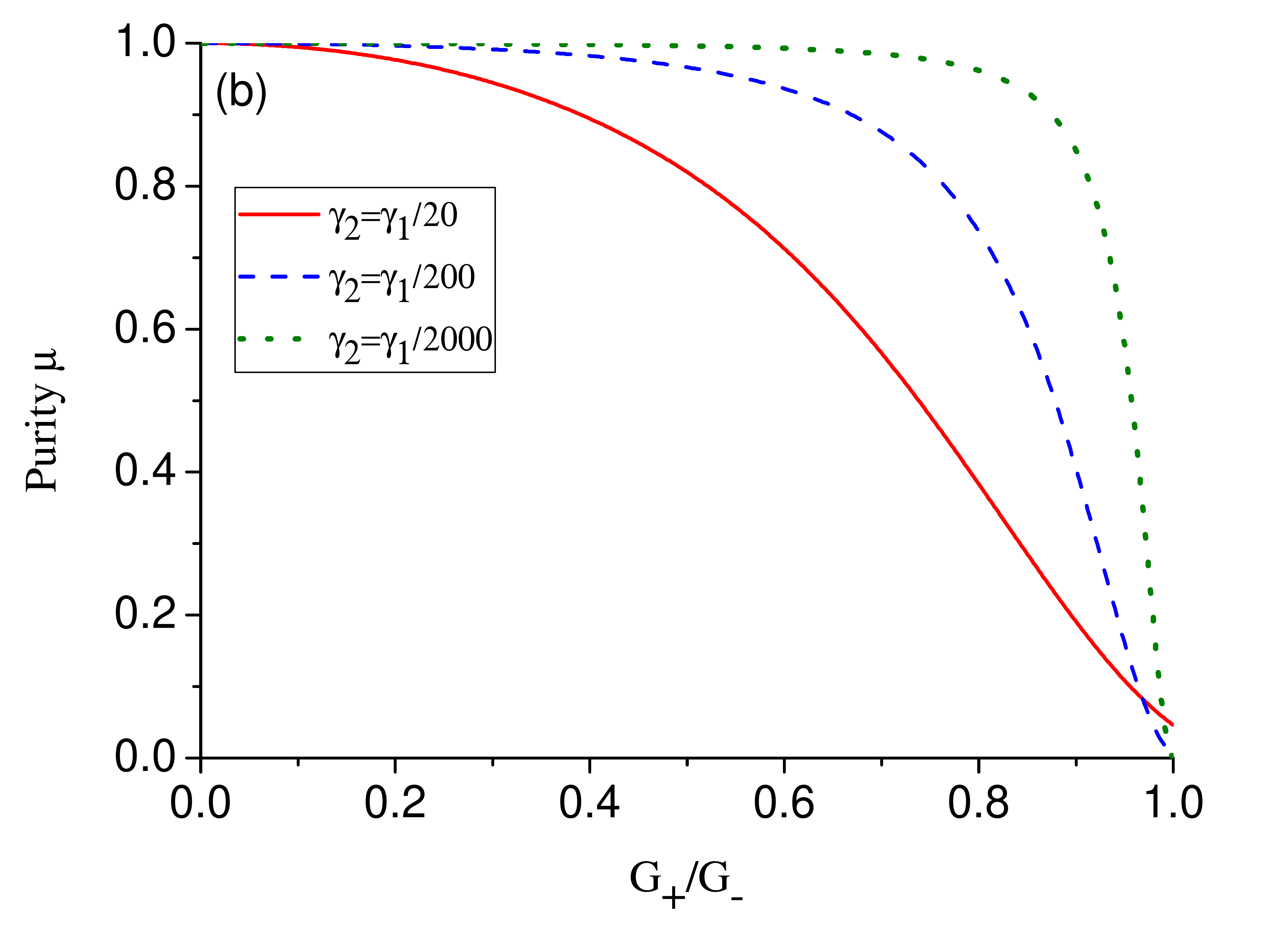}\label{fig:PB}}
   \subfigure{\includegraphics[width=0.42\textwidth]{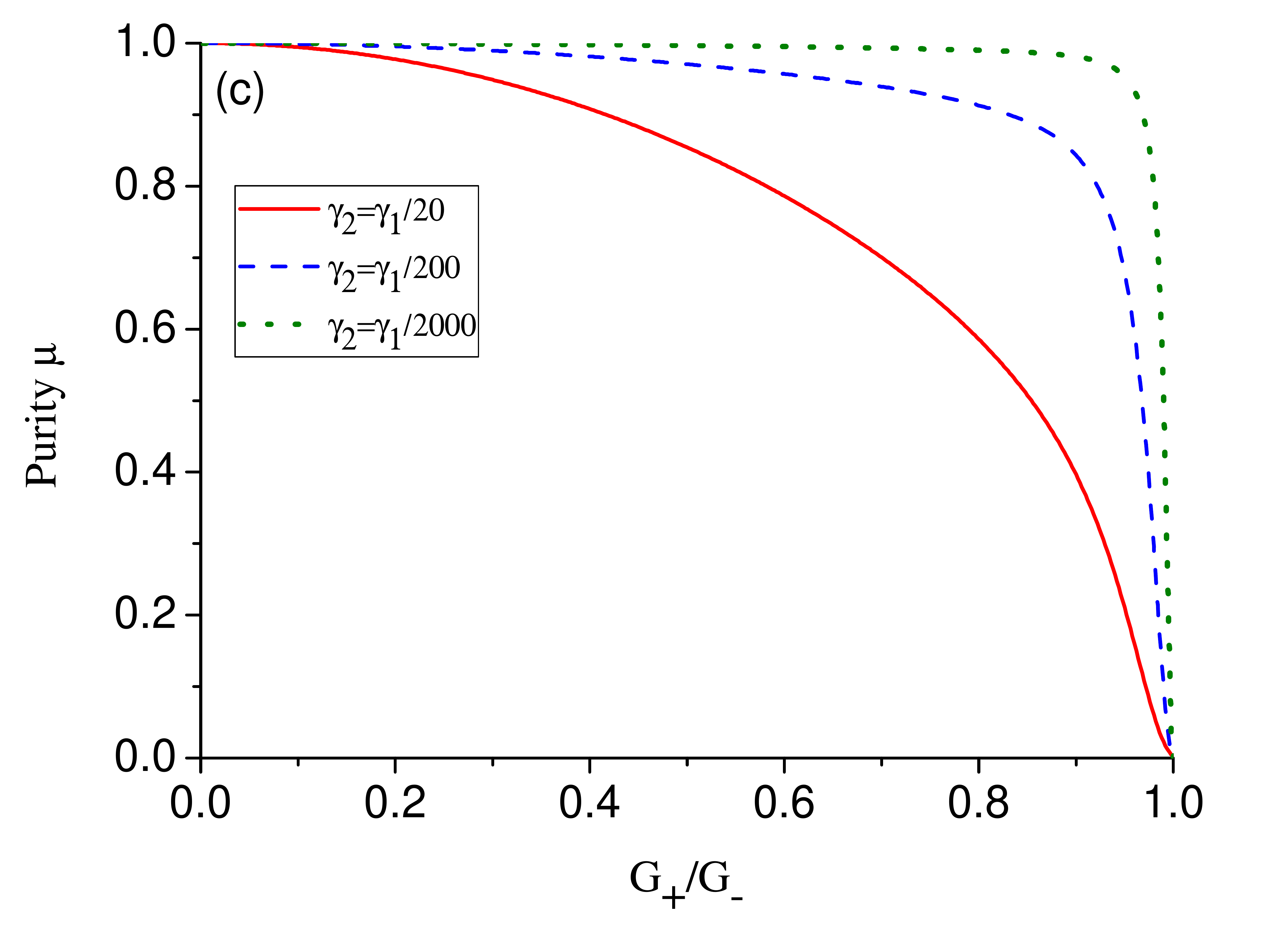}\label{fig:PC}}
   \caption{\label{fig:Purity}
   (Color online) Steady-state purity of the cavity mode $d$ and the mechanical mode $b_2$ against the ratio of the effective couplings $G_+/G_-$ for different values of $\delta$. (a) $\delta =10\gamma_1$, (b) $\delta =5\gamma_1$, and (c) $\delta =\gamma_1$. All other parameters are the same as those in Fig.~\ref{fig:EVC}.}
\end{figure}

We next study the steady-state entanglement ($\dot{\sigma}(t)=0$ in the stationary limit $t\gg1/\kappa,\gamma_{1,2}$ if the system is stable) with the time-independent Hamiltonian in Eqs.~(\ref{HRWA}) and ~(\ref{HRWA1}) under the rotating-wave approximation (by dropping all time-dependent terms in Eq.~(\ref{M})). Fig.~\ref{fig:EVC} displays the steady-state entanglement $E_N$ of the cavity mode $d$ and the mechanical mode $b_2$ as functions of the coupling asymmetry $G_+/G_-$ for different $\delta$ with zero bath occupations for all modes, where the downward triangle denotes the optimal value of each curve.  Apparently, $E_N$ is a non-monotonic function of $G_+/G_-$ in any given set of parameters and takes a maximum for a specific $G_+/G_-$. The phenomenon is similar to that in Refs.~\cite{wang2013reservoir,woolley2014two,chen2015dissipation}, and it can be explained as follows. The relation $\tanh r=G_+/G_-$ indicates that the increase of the ratio $G_+/G_-$ can raise the squeezing parameter $r$, which is beneficial for enhancing the entanglement. But, from another point of view, the increase in $G_+/G_-$ (with $G_-$ fixed) accompanies the decline of  effective coupling $G=\sqrt{G_-^2-G_+^2}$ between the sum mode $\beta_{sum}$ and the mechanical mode $b_1$ , which is harmful for the cooling effect and thus reduces the amount of entanglement. The best value is obtained when the two competing effects balance. In addition, we find that the smaller the ratio $\gamma_2/\gamma_1$, the lager the maximal entanglement $E_N$ and the optimal $G_+/G_-$ in each figure. Since the entanglement generation is largely based on cooling the Bogoliubov modes via the dissipative dynamics of the mechanical mode $b_1$, one would expect that a strong damping rate $\gamma_1$ of $b_1$ and simultaneously weak damping rates $\gamma_2$ of $b_2$ and  $\kappa$ of $d$ should increase the peak entanglement $E_N$ (corresponding to bigger $G_+/G_-$). Comparing Figs.~\ref{fig:EVCA}, \ref{fig:EVCB} and \ref{fig:EVCC} with different values of $\delta$, one can find that the achievable entanglement is also dependent on $\delta$, which is the  effective coupling between the sum mode $\beta_{sum}$ and the difference mode $\beta_{diff}$ and induces the cooling process of $\beta_{diff}$.

\begin{figure}[t]
  {\includegraphics[width=0.42\textwidth]{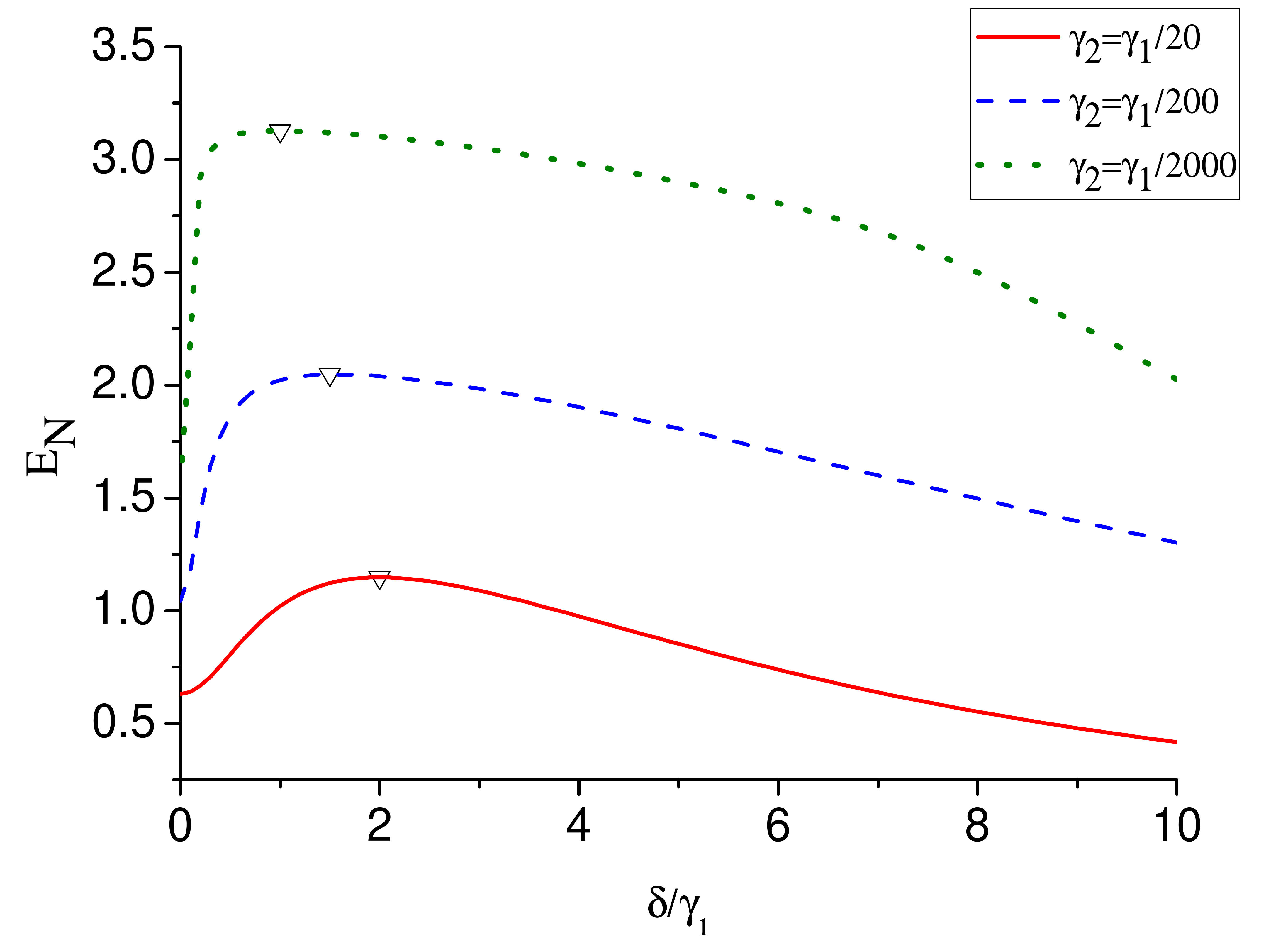}}
  \caption{\label{fig:EVD}
  (Color online) Stationary cavity-mechanical entanglement $E_N$ versus the effective coupling $\delta$. The sets of parameters corresponding to different lines are $\gamma_2=\gamma_1/20,G_+/G_-=0.604$ (red solid); $\gamma_2=\gamma_1/200,G_+/G_-=0.786$ (blue dashed); and $\gamma_2=\gamma_1/2000,G_+/G_-=0.918$ (olive dotted). The other parameters are: $G_-=2.5\gamma_1$, $\kappa=\gamma_2$, $\bar{n}_1=\bar{n}_2=\bar{n}_d=0$. }
\end{figure}

\begin{figure}[t]
  {\includegraphics[width=0.45\textwidth]{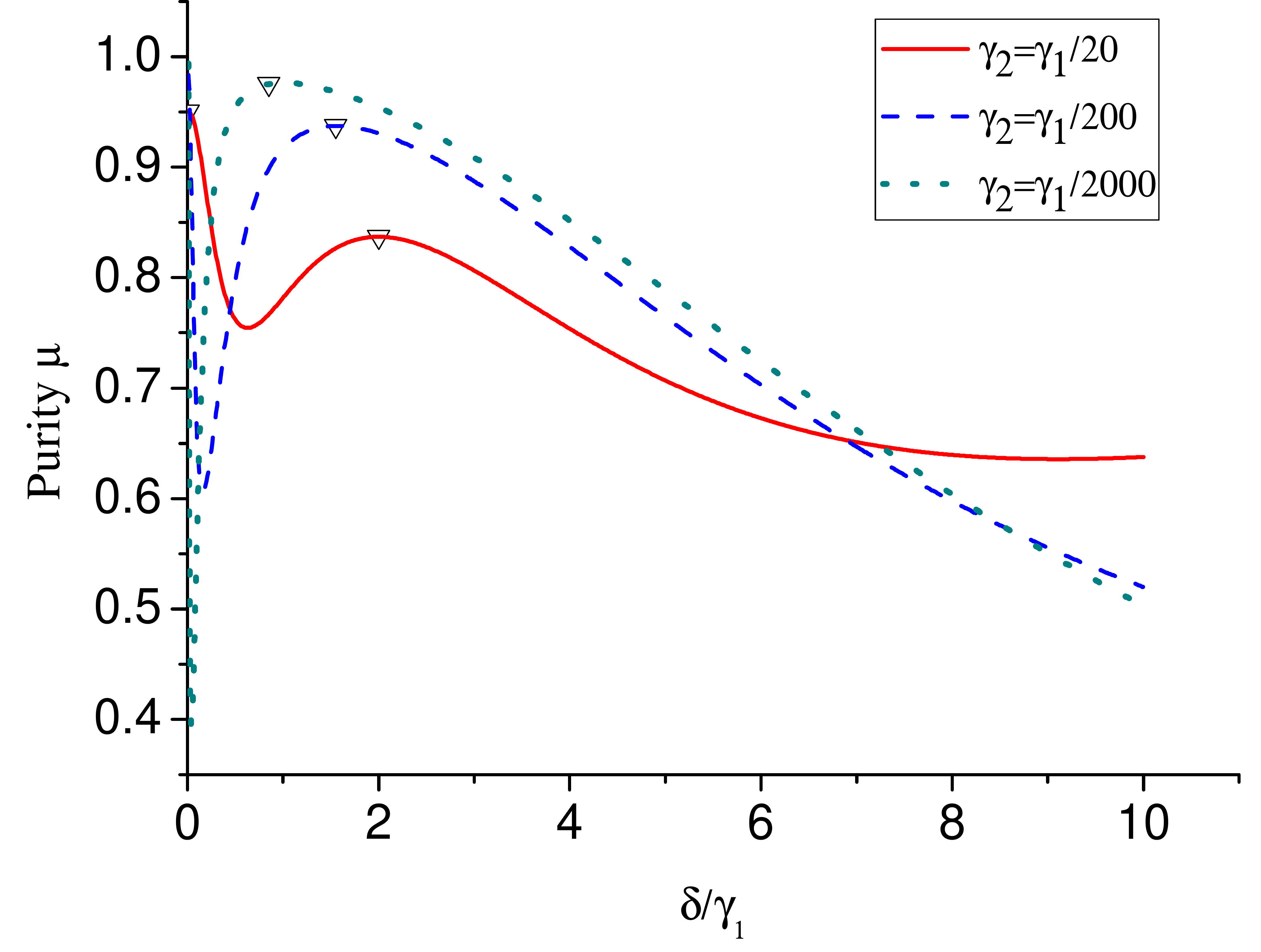}}
  \caption{\label{fig:PVD}
  (Color online) Steady-state purity of the cavity mode $d$ and the mechanical mode $b_2$ versus the effective coupling $\delta$. All parameters are the same as those in Fig.~\ref{fig:EVD}.}
\end{figure}

Fig.~\ref{fig:Purity} shows the purity as functions of the coupling asymmetry $G_+/G_-$. Clearly, we can observe that the purity is inversely correlated to $G_+/G_-$. If $\gamma_2$ is small enough compared to $\gamma_1$, one can keep the high purity ($\approx1$) of the steady states over a wide range of $G_+/G_-$. However, in order to enhance the entanglement, one needs a larger squeezing parameter $r=\tanh^{-1}(G_+/G_-)$ (i.e. larger $G_+/G_-$) which, on the other hand, weakens the effective coupling $G=\sqrt{G_-^2-G_+^2}$ and, hence, cripples the cooling process of Bogoliubov modes toward a pure ground state via the dissipation of $b_1$. For the sake of gaining a large amount of entanglement while retaining the relatively high purity of the  entangled states, we can select proper detuning $\delta$ as shown in Figs.~\ref{fig:EVD} and \ref{fig:PVD}, where the downward triangles indicate the optimal values of the corresponding curves. Note that the chosen coupling asymmetry $G_+/G_-$  for each $\gamma_2$ is the value where $E_N$ takes the maximum in Fig.~\ref{fig:EVCA}. Remarkably, one can find specific $\delta$ where both the entanglement and the purity take the local maximum. For example, when  $\gamma_2=\gamma_1/2000,G_+/G_-=0.918,\delta\approx\gamma_1$, we have $E_N\approx3.2$ and $\mu\approx0.98$. In other words, our scheme allows the generation of  highly pure and strongly entangled optomechanical states.

\begin{figure}
   \centering
   \subfigure{\includegraphics[width=0.42\textwidth]{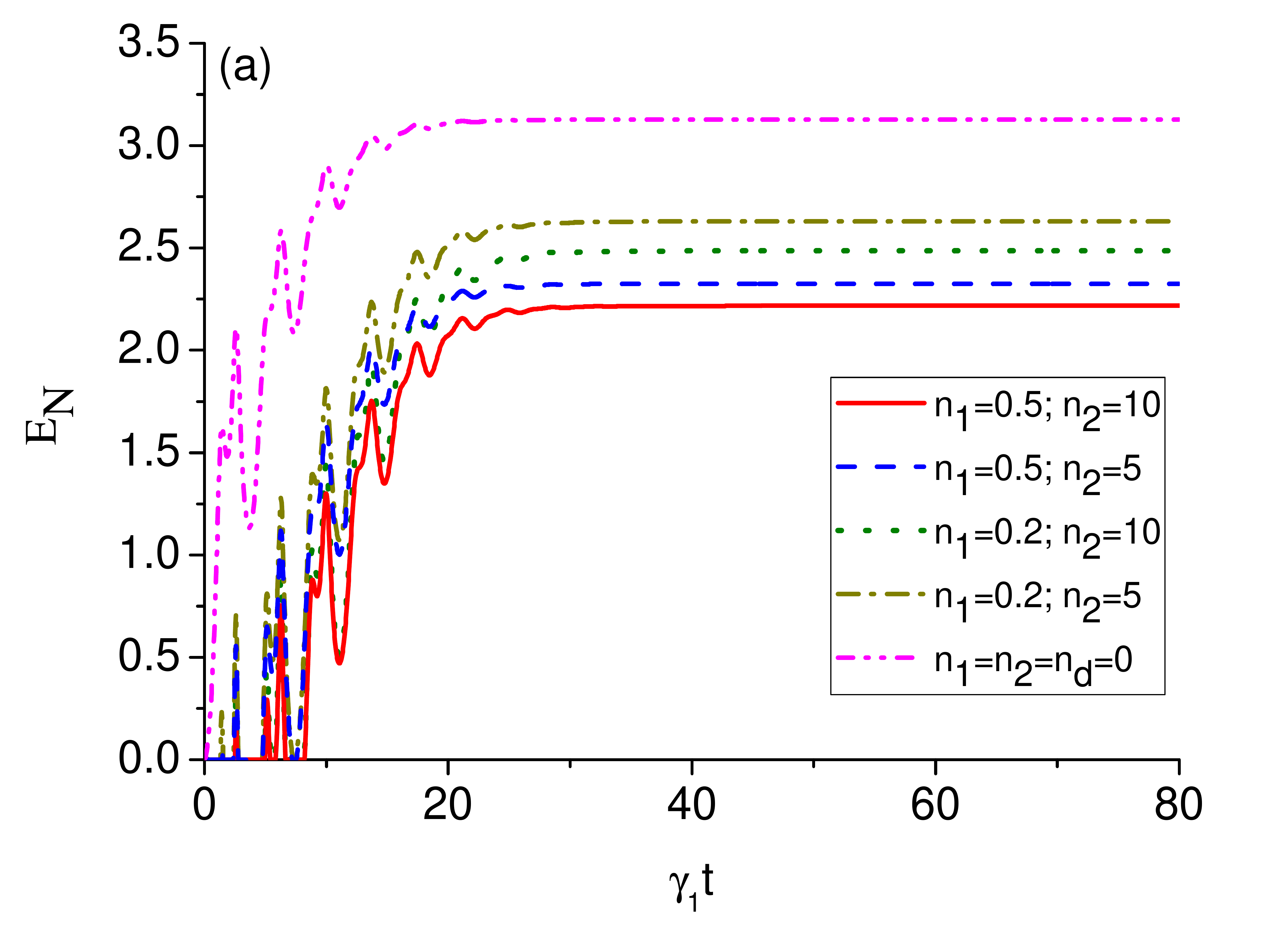}\label{fig:ETA}}
   \subfigure{\includegraphics[width=0.42\textwidth]{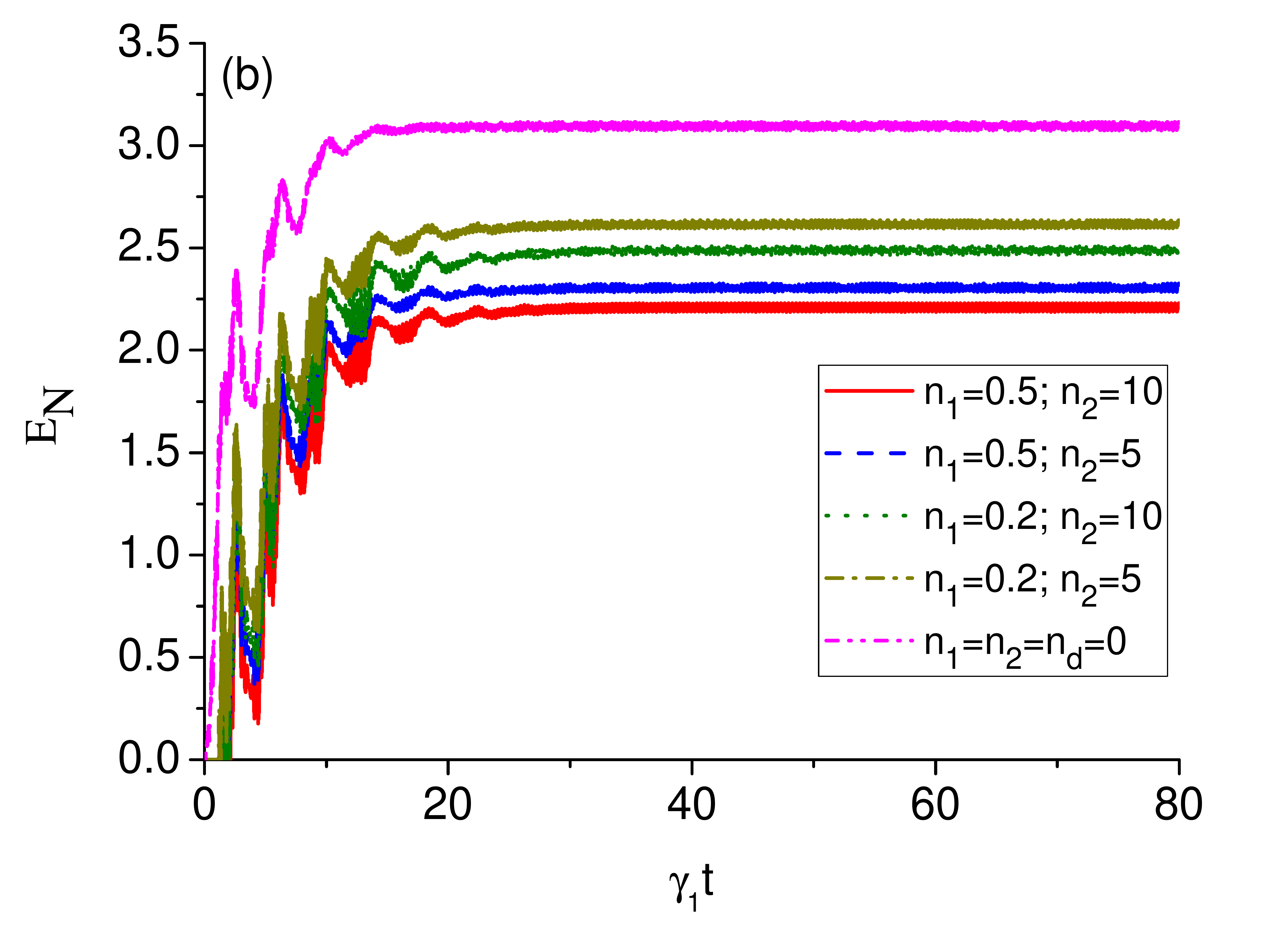}\label{fig:ETB}}
   \subfigure{\includegraphics[width=0.42\textwidth]{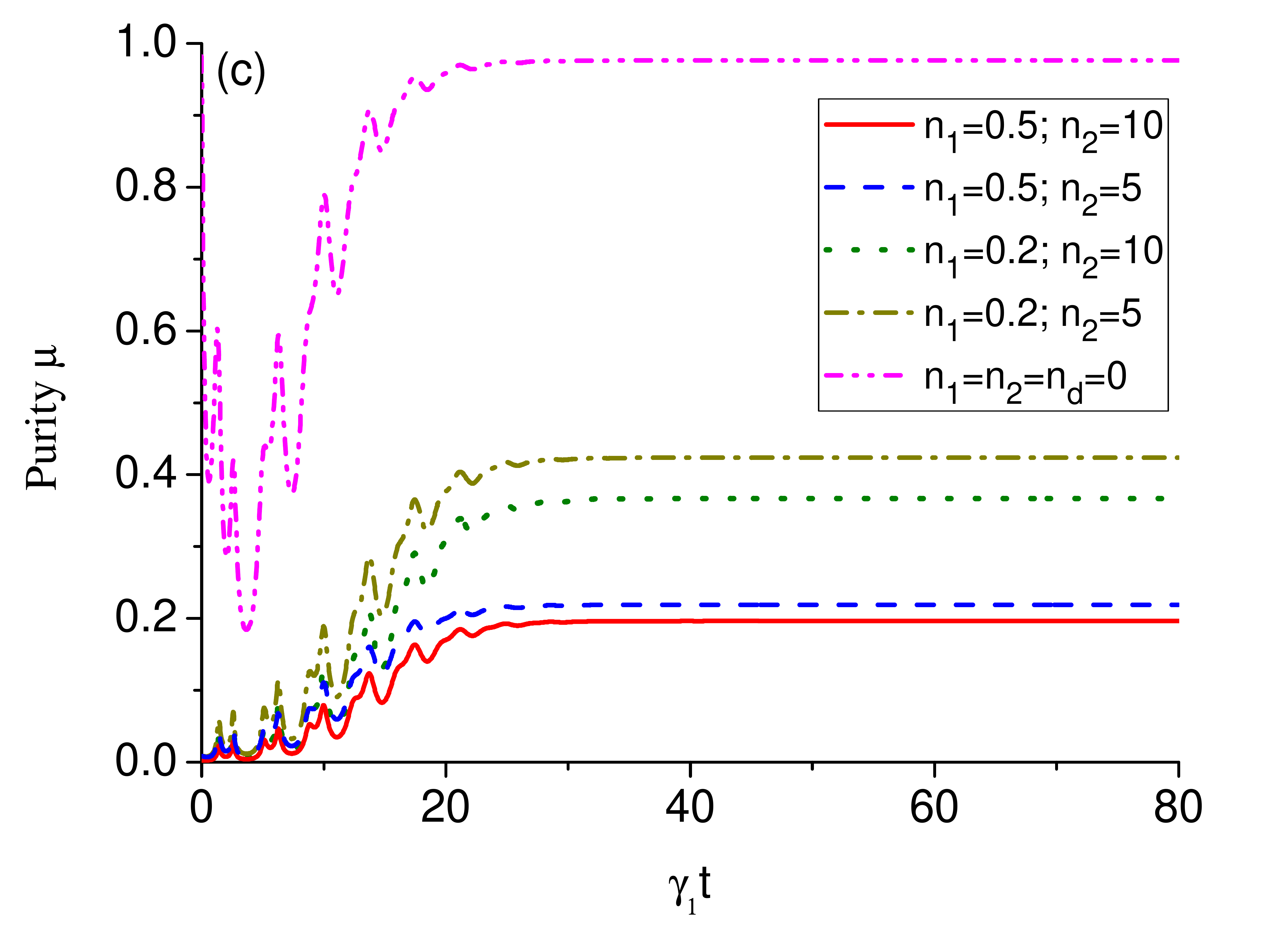}\label{fig:PTA}}
   \subfigure{\includegraphics[width=0.42\textwidth]{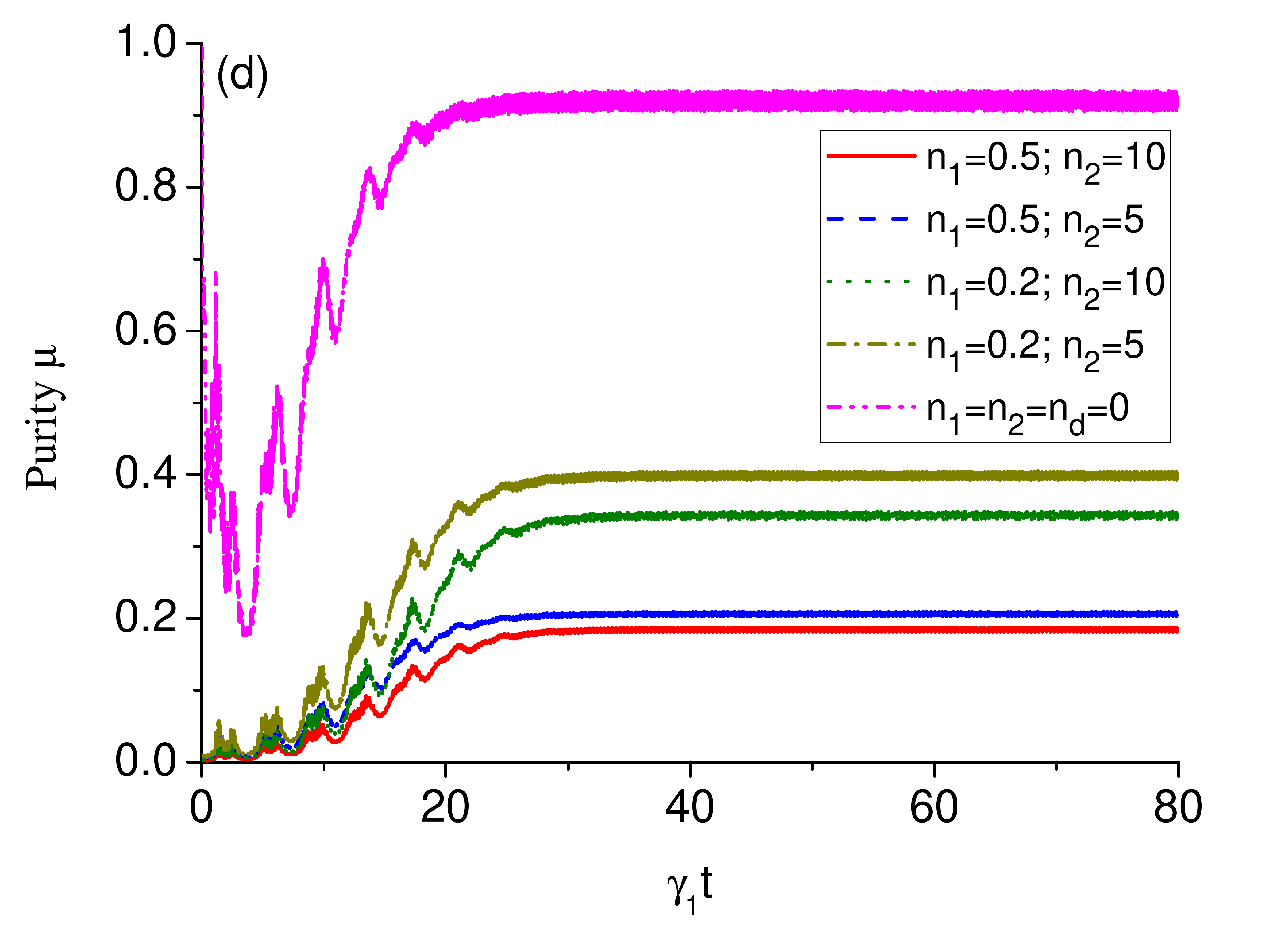}\label{fig:PTB}}
   \caption{\label{fig:EPT}
   (Color online) The time evolution of the entanglement [(a) and (b)] and purity [(c) and (d)] of the quantum states of the cavity mode $d$ and mechanical mode $b_2$ with [(b) and (d)] and without [(a) and (c)] the non-resonant terms. The parameters are $G_-=2.5\gamma_1$, $G_+=0.918G_-$, $\kappa=\gamma_2=\gamma_1/2000$, $\delta=\gamma_1$, $\bar{n}_2=\bar{n}_d$, $\omega_1=10\gamma_1$, and $\omega_2=100\gamma_1$.}
\end{figure}

To find the optimal $\delta_{opt}$, one can recall the  Hamiltonian under the rotating-wave approximation in Eq.~(\ref{HRWA1}). The sum mode $\beta_{sum}$ is simultaneously  coupled to  the difference mode $\beta_{diff}$ and the mechanical mode $b_1$ with  beam-splitter-like coupling strengths $\delta$ and $\sqrt{2}G$ respectively. The coupling between $\beta_{sum}$ and $b_1$ induces the cooling process of $\beta_{sum}$, while the coupling between  $\beta_{sum}$ and $\beta_{diff}$  is responsible for cooling the $\beta_{diff}$ mode. For a given (fixed) set of parameters $G_+$, $G_-$, on the one hand, if $\delta$ is too small (relative to $G=\sqrt{G_-^{2}-G_+^2}$), $\beta_{diff}$ can not be effectively cooled by $\beta_{sum}$. For example, when $\delta$  approaches 0, only the $\beta_{sum}$ mode can be cooled by
$b_1$. On the other hand, if $\delta$ is too large, i.e., $\beta_{diff}$ and $\beta_{sum}$ are strongly coupled, the quanta are confined  and  swap rapidly between them. Hence, $\beta_{sum}$ can not be effectively cooled by $b_1$ in this case.  For different sets of parameters $G_+$ and $G_-$, one would expect some  moderate values of $\delta$ that  correspond to maximum entanglement and  purity. In fact, we have found  that the optimal $\delta_{opt}$ is approximately equal to $G$ from Figs.~\ref{fig:EVD} and  \ref{fig:PVD}, where $\delta_{opt}\approx G\approx2\gamma_1$ for red solid lines,  $\delta_{opt}\approx G\approx 1.5\gamma_1$  for blue dashed lines, and $\delta_{opt}\approx G\approx 0.99\gamma_1$ for olive dotted lines.

So far all of our discussions have been restricted to the rotating-wave approximation. To study the effects of non-resonant terms  of the linearized Hamiltonian in Eq.~(\ref{Hlin}), we plot  in Fig.~\ref{fig:EPT} the time evolution of the entanglement and purity with (Figs.~\ref{fig:ETB} and~\ref{fig:PTB}) and without (Figs.~\ref{fig:ETA} and~\ref{fig:PTA}) the non-resonant terms for some bath occupancies. We study the system dynamics  by numerically solving  the differential equation of the covariance matrix in Eq.~(\ref{dotsigma}) with the initial states of  all modes assumed to be in thermal equilibrium with their local baths.
When performing the numerical simulations, the effects of non-resonant terms are included by using the full time-dependent coefficient matrix $M(t)$ in Eq.~(\ref{M}) containing all time-dependent terms. We find that the non-resonant terms only induce small oscillations and do not significantly reduce the amount of steady-state entanglement and purity in the long-time limit, suggesting that the rotating-wave approximation is indeed valid.

\section{conclusions}

In summary, we have proposed an effective approach to generate pure and strong steady-state opto-mechanical entanglement (or optical-microwave entanglement) in a hybrid modulated three-mode optomechanical system. By applying a proper two-tone driving of the cavity and modulating coupling strength between two mechanical oscillators (or between a mechanical oscillator and a superconducting transmission line resonator), one can prepare the two target modes of the system in an entangled steady state. The proposal uses an intermediate mechanical mode acting as an engineered reservoir to effectively cool both Bogoliubov modes of the target modes to near their ground state via the beam-splitter-like interactions. Our approach allows the generation of a highly pure and strongly entangled steady state, by properly choosing not only the ratio of the effective optomechanical couplings but also the cavity-pump detuning.

\section*{Acknowledgments}
C. G. Liao, H. Xie and X. M. Lin are supported by the National Natural Science Foundation of China (Grants No.~61275215 and No.~11674059), the Natural Science Foundation of Fujian Province of China (Grants No.~2016J01009 and No.~2013J01008), the Educational Committee of Fujian Province of China (Grants No.~JAT160687 and No.~JA14397), the 2016 Annual College Funds for Distinguished Young Scientists of Fujian Province of China, and funds from Fujian Polytechnic of Information Technology (Grant No.~Y17104).
R. X. Chen is supported by the Office of Naval Research (Award No. N00014-16-1-3054) and Robert A. Welch Foundation (Grant No. A-1261).
\bibliography{entanglement}

\begin{thebibliography}{56}%
\makeatletter
\providecommand \@ifxundefined [1]{%
 \@ifx{#1\undefined}
}%
\providecommand \@ifnum [1]{%
 \ifnum #1\expandafter \@firstoftwo
 \else \expandafter \@secondoftwo
 \fi
}%
\providecommand \@ifx [1]{%
 \ifx #1\expandafter \@firstoftwo
 \else \expandafter \@secondoftwo
 \fi
}%
\providecommand \natexlab [1]{#1}%
\providecommand \enquote  [1]{``#1''}%
\providecommand \bibnamefont  [1]{#1}%
\providecommand \bibfnamefont [1]{#1}%
\providecommand \citenamefont [1]{#1}%
\providecommand \href@noop [0]{\@secondoftwo}%
\providecommand \href [0]{\begingroup \@sanitize@url \@href}%
\providecommand \@href[1]{\@@startlink{#1}\@@href}%
\providecommand \@@href[1]{\endgroup#1\@@endlink}%
\providecommand \@sanitize@url [0]{\catcode `\\12\catcode `\$12\catcode
  `\&12\catcode `\#12\catcode `\^12\catcode `\_12\catcode `\%12\relax}%
\providecommand \@@startlink[1]{}%
\providecommand \@@endlink[0]{}%
\providecommand \url  [0]{\begingroup\@sanitize@url \@url }%
\providecommand \@url [1]{\endgroup\@href {#1}{\urlprefix }}%
\providecommand \urlprefix  [0]{URL }%
\providecommand \Eprint [0]{\href }%
\providecommand \doibase [0]{http://dx.doi.org/}%
\providecommand \selectlanguage [0]{\@gobble}%
\providecommand \bibinfo  [0]{\@secondoftwo}%
\providecommand \bibfield  [0]{\@secondoftwo}%
\providecommand \translation [1]{[#1]}%
\providecommand \BibitemOpen [0]{}%
\providecommand \bibitemStop [0]{}%
\providecommand \bibitemNoStop [0]{.\EOS\space}%
\providecommand \EOS [0]{\spacefactor3000\relax}%
\providecommand \BibitemShut  [1]{\csname bibitem#1\endcsname}%
\let\auto@bib@innerbib\@empty
\bibitem [{\citenamefont {Fabre}\ \emph {et~al.}(1994)\citenamefont {Fabre},
  \citenamefont {Pinard}, \citenamefont {Bourzeix}, \citenamefont {Heidmann},
  \citenamefont {Giacobino},\ and\ \citenamefont
  {Reynaud}}]{fabre1994quantum-noise}%
  \BibitemOpen
  \bibfield  {author} {\bibinfo {author} {\bibfnamefont {C.}~\bibnamefont
  {Fabre}}, \bibinfo {author} {\bibfnamefont {M.}~\bibnamefont {Pinard}},
  \bibinfo {author} {\bibfnamefont {S.}~\bibnamefont {Bourzeix}}, \bibinfo
  {author} {\bibfnamefont {A.}~\bibnamefont {Heidmann}}, \bibinfo {author}
  {\bibfnamefont {E.}~\bibnamefont {Giacobino}}, \ and\ \bibinfo {author}
  {\bibfnamefont {S.}~\bibnamefont {Reynaud}},\ }\href {\doibase
  10.1103/PhysRevA.49.1337} {\bibfield  {journal} {\bibinfo  {journal} {Phys.
  Rev. A}\ }\textbf {\bibinfo {volume} {49}},\ \bibinfo {pages} {1337}
  (\bibinfo {year} {1994})}\BibitemShut {NoStop}%
\bibitem [{\citenamefont {Mancini}\ and\ \citenamefont
  {Tombesi}(1994)}]{mancini1994quantum}%
  \BibitemOpen
  \bibfield  {author} {\bibinfo {author} {\bibfnamefont {S.}~\bibnamefont
  {Mancini}}\ and\ \bibinfo {author} {\bibfnamefont {P.}~\bibnamefont
  {Tombesi}},\ }\href {\doibase 10.1103/PhysRevA.49.4055} {\bibfield  {journal}
  {\bibinfo  {journal} {Phys. Rev. A}\ }\textbf {\bibinfo {volume} {49}},\
  \bibinfo {pages} {4055} (\bibinfo {year} {1994})}\BibitemShut {NoStop}%
\bibitem [{\citenamefont {Jacobs}\ \emph {et~al.}(1994)\citenamefont {Jacobs},
  \citenamefont {Tombesi}, \citenamefont {Collett},\ and\ \citenamefont
  {Walls}}]{Jacobs1994Quantum}%
  \BibitemOpen
  \bibfield  {author} {\bibinfo {author} {\bibfnamefont {K.}~\bibnamefont
  {Jacobs}}, \bibinfo {author} {\bibfnamefont {P.}~\bibnamefont {Tombesi}},
  \bibinfo {author} {\bibfnamefont {M.~J.}\ \bibnamefont {Collett}}, \ and\
  \bibinfo {author} {\bibfnamefont {D.~F.}\ \bibnamefont {Walls}},\ }\href
  {\doibase 10.1103/PhysRevA.49.1961} {\bibfield  {journal} {\bibinfo
  {journal} {Phys. Rev. A}\ }\textbf {\bibinfo {volume} {49}},\ \bibinfo
  {pages} {1961} (\bibinfo {year} {1994})}\BibitemShut {NoStop}%
\bibitem [{\citenamefont {Pinard}\ \emph {et~al.}(1995)\citenamefont {Pinard},
  \citenamefont {Fabre},\ and\ \citenamefont {Heidmann}}]{Pinard1995Quantum}%
  \BibitemOpen
  \bibfield  {author} {\bibinfo {author} {\bibfnamefont {M.}~\bibnamefont
  {Pinard}}, \bibinfo {author} {\bibfnamefont {C.}~\bibnamefont {Fabre}}, \
  and\ \bibinfo {author} {\bibfnamefont {A.}~\bibnamefont {Heidmann}},\ }\href
  {\doibase 10.1103/PhysRevA.51.2443} {\bibfield  {journal} {\bibinfo
  {journal} {Phys. Rev. A}\ }\textbf {\bibinfo {volume} {51}},\ \bibinfo
  {pages} {2443} (\bibinfo {year} {1995})}\BibitemShut {NoStop}%
\bibitem [{\citenamefont {Bose}\ \emph {et~al.}(1997)\citenamefont {Bose},
  \citenamefont {Jacobs},\ and\ \citenamefont {Knight}}]{Bose1997Preparation}%
  \BibitemOpen
  \bibfield  {author} {\bibinfo {author} {\bibfnamefont {S.}~\bibnamefont
  {Bose}}, \bibinfo {author} {\bibfnamefont {K.}~\bibnamefont {Jacobs}}, \ and\
  \bibinfo {author} {\bibfnamefont {P.~L.}\ \bibnamefont {Knight}},\ }\href
  {\doibase 10.1103/PhysRevA.56.4175} {\bibfield  {journal} {\bibinfo
  {journal} {Phys. Rev. A}\ }\textbf {\bibinfo {volume} {56}},\ \bibinfo
  {pages} {4175} (\bibinfo {year} {1997})}\BibitemShut {NoStop}%
\bibitem [{\citenamefont {Mancini}\ \emph {et~al.}(1997)\citenamefont
  {Mancini}, \citenamefont {Man'ko},\ and\ \citenamefont
  {Tombesi}}]{Mancini1996Ponderomotive}%
  \BibitemOpen
  \bibfield  {author} {\bibinfo {author} {\bibfnamefont {S.}~\bibnamefont
  {Mancini}}, \bibinfo {author} {\bibfnamefont {V.~I.}\ \bibnamefont {Man'ko}},
  \ and\ \bibinfo {author} {\bibfnamefont {P.}~\bibnamefont {Tombesi}},\ }\href
  {\doibase 10.1103/PhysRevA.55.3042} {\bibfield  {journal} {\bibinfo
  {journal} {Phys. Rev. A}\ }\textbf {\bibinfo {volume} {55}},\ \bibinfo
  {pages} {3042} (\bibinfo {year} {1997})}\BibitemShut {NoStop}%
\bibitem [{\citenamefont {Schwab}\ and\ \citenamefont
  {Roukes}(2005)}]{Schwab2005Putting}%
  \BibitemOpen
  \bibfield  {author} {\bibinfo {author} {\bibfnamefont {K.~C.}\ \bibnamefont
  {Schwab}}\ and\ \bibinfo {author} {\bibfnamefont {M.~L.}\ \bibnamefont
  {Roukes}},\ }\href {http://physicstoday.scitation.org/doi/10.1063/1.2012461}
  {\bibfield  {journal} {\bibinfo  {journal} {Phys. Today}\ }\textbf {\bibinfo
  {volume} {58}},\ \bibinfo {pages} {36} (\bibinfo {year} {2005})}\BibitemShut
  {NoStop}%
\bibitem [{\citenamefont {Cohadon}\ \emph {et~al.}(1999)\citenamefont
  {Cohadon}, \citenamefont {Heidmann},\ and\ \citenamefont
  {Pinard}}]{Cohadon1999Cooling}%
  \BibitemOpen
  \bibfield  {author} {\bibinfo {author} {\bibfnamefont {P.~F.}\ \bibnamefont
  {Cohadon}}, \bibinfo {author} {\bibfnamefont {A.}~\bibnamefont {Heidmann}}, \
  and\ \bibinfo {author} {\bibfnamefont {M.}~\bibnamefont {Pinard}},\ }\href
  {\doibase 10.1103/PhysRevLett.83.3174} {\bibfield  {journal} {\bibinfo
  {journal} {Phys. Rev. Lett.}\ }\textbf {\bibinfo {volume} {83}},\ \bibinfo
  {pages} {3174} (\bibinfo {year} {1999})}\BibitemShut {NoStop}%
\bibitem [{\citenamefont {Chen}(2013)}]{Chen2013Macroscopic}%
  \BibitemOpen
  \bibfield  {author} {\bibinfo {author} {\bibfnamefont {Y.}~\bibnamefont
  {Chen}},\ }\href {http://stacks.iop.org/0953-4075/46/i=10/a=104001}
  {\bibfield  {journal} {\bibinfo  {journal} {J. Phys. B}\ }\textbf {\bibinfo
  {volume} {46}},\ \bibinfo {pages} {104001} (\bibinfo {year}
  {2013})}\BibitemShut {NoStop}%
\bibitem [{\citenamefont {Hammerer}\ \emph {et~al.}(2009)\citenamefont
  {Hammerer}, \citenamefont {Wallquist}, \citenamefont {Genes}, \citenamefont
  {Ludwig}, \citenamefont {Marquardt}, \citenamefont {Treutlein}, \citenamefont
  {Zoller}, \citenamefont {Ye},\ and\ \citenamefont
  {Kimble}}]{Hammerer2009Strong}%
  \BibitemOpen
  \bibfield  {author} {\bibinfo {author} {\bibfnamefont {K.}~\bibnamefont
  {Hammerer}}, \bibinfo {author} {\bibfnamefont {M.}~\bibnamefont {Wallquist}},
  \bibinfo {author} {\bibfnamefont {C.}~\bibnamefont {Genes}}, \bibinfo
  {author} {\bibfnamefont {M.}~\bibnamefont {Ludwig}}, \bibinfo {author}
  {\bibfnamefont {F.}~\bibnamefont {Marquardt}}, \bibinfo {author}
  {\bibfnamefont {P.}~\bibnamefont {Treutlein}}, \bibinfo {author}
  {\bibfnamefont {P.}~\bibnamefont {Zoller}}, \bibinfo {author} {\bibfnamefont
  {J.}~\bibnamefont {Ye}}, \ and\ \bibinfo {author} {\bibfnamefont {H.~J.}\
  \bibnamefont {Kimble}},\ }\href {\doibase 10.1103/PhysRevLett.103.063005}
  {\bibfield  {journal} {\bibinfo  {journal} {Phys. Rev. Lett.}\ }\textbf
  {\bibinfo {volume} {103}},\ \bibinfo {pages} {063005} (\bibinfo {year}
  {2009})}\BibitemShut {NoStop}%
\bibitem [{\citenamefont {Hammerer}\ \emph {et~al.}(2010)\citenamefont
  {Hammerer}, \citenamefont {S\o{}rensen},\ and\ \citenamefont
  {Polzik}}]{Polzik2008Quantum}%
  \BibitemOpen
  \bibfield  {author} {\bibinfo {author} {\bibfnamefont {K.}~\bibnamefont
  {Hammerer}}, \bibinfo {author} {\bibfnamefont {A.~S.}\ \bibnamefont
  {S\o{}rensen}}, \ and\ \bibinfo {author} {\bibfnamefont {E.~S.}\ \bibnamefont
  {Polzik}},\ }\href {\doibase 10.1103/RevModPhys.82.1041} {\bibfield
  {journal} {\bibinfo  {journal} {Rev. Mod. Phys.}\ }\textbf {\bibinfo {volume}
  {82}},\ \bibinfo {pages} {1041} (\bibinfo {year} {2010})}\BibitemShut
  {NoStop}%
\bibitem [{\citenamefont {Chen}\ \emph {et~al.}(2010)\citenamefont {Chen},
  \citenamefont {Goldbaum}, \citenamefont {Bhattacharya},\ and\ \citenamefont
  {Meystre}}]{Chen2010Classical}%
  \BibitemOpen
  \bibfield  {author} {\bibinfo {author} {\bibfnamefont {W.}~\bibnamefont
  {Chen}}, \bibinfo {author} {\bibfnamefont {D.~S.}\ \bibnamefont {Goldbaum}},
  \bibinfo {author} {\bibfnamefont {M.}~\bibnamefont {Bhattacharya}}, \ and\
  \bibinfo {author} {\bibfnamefont {P.}~\bibnamefont {Meystre}},\ }\href
  {\doibase 10.1103/PhysRevA.81.053833} {\bibfield  {journal} {\bibinfo
  {journal} {Phys. Rev. A}\ }\textbf {\bibinfo {volume} {81}},\ \bibinfo
  {pages} {053833} (\bibinfo {year} {2010})}\BibitemShut {NoStop}%
\bibitem [{\citenamefont {Jing}\ \emph {et~al.}(2011)\citenamefont {Jing},
  \citenamefont {Goldbaum}, \citenamefont {Buchmann},\ and\ \citenamefont
  {Meystre}}]{Jing2011Quantum}%
  \BibitemOpen
  \bibfield  {author} {\bibinfo {author} {\bibfnamefont {H.}~\bibnamefont
  {Jing}}, \bibinfo {author} {\bibfnamefont {D.~S.}\ \bibnamefont {Goldbaum}},
  \bibinfo {author} {\bibfnamefont {L.}~\bibnamefont {Buchmann}}, \ and\
  \bibinfo {author} {\bibfnamefont {P.}~\bibnamefont {Meystre}},\ }\href
  {\doibase 10.1103/PhysRevLett.106.223601} {\bibfield  {journal} {\bibinfo
  {journal} {Phys. Rev. Lett.}\ }\textbf {\bibinfo {volume} {106}},\ \bibinfo
  {pages} {223601} (\bibinfo {year} {2011})}\BibitemShut {NoStop}%
\bibitem [{\citenamefont {O'Connell}\ \emph {et~al.}(2010)\citenamefont
  {O'Connell}, \citenamefont {Hofheinz}, \citenamefont {Ansmann}, \citenamefont
  {Bialczak}, \citenamefont {Lenander}, \citenamefont {Lucero}, \citenamefont
  {Neeley}, \citenamefont {Sank}, \citenamefont {Wang}, \citenamefont {Weides},
  \citenamefont {Wenner}, \citenamefont {Martinis},\ and\ \citenamefont
  {Cleland}}]{O2010Quantum}%
  \BibitemOpen
  \bibfield  {author} {\bibinfo {author} {\bibfnamefont {A.~D.}\ \bibnamefont
  {O'Connell}}, \bibinfo {author} {\bibfnamefont {M.}~\bibnamefont {Hofheinz}},
  \bibinfo {author} {\bibfnamefont {M.}~\bibnamefont {Ansmann}}, \bibinfo
  {author} {\bibfnamefont {R.~C.}\ \bibnamefont {Bialczak}}, \bibinfo {author}
  {\bibfnamefont {M.}~\bibnamefont {Lenander}}, \bibinfo {author}
  {\bibfnamefont {E.}~\bibnamefont {Lucero}}, \bibinfo {author} {\bibfnamefont
  {M.}~\bibnamefont {Neeley}}, \bibinfo {author} {\bibfnamefont
  {D.}~\bibnamefont {Sank}}, \bibinfo {author} {\bibfnamefont {H.}~\bibnamefont
  {Wang}}, \bibinfo {author} {\bibfnamefont {M.}~\bibnamefont {Weides}},
  \bibinfo {author} {\bibfnamefont {J.}~\bibnamefont {Wenner}}, \bibinfo
  {author} {\bibfnamefont {J.~M.}\ \bibnamefont {Martinis}}, \ and\ \bibinfo
  {author} {\bibfnamefont {A.~N.}\ \bibnamefont {Cleland}},\ }\href
  {http://dx.doi.org/10.1038/nature08967} {\bibfield  {journal} {\bibinfo
  {journal} {Nature (London)}\ }\textbf {\bibinfo {volume} {464}},\ \bibinfo
  {pages} {697} (\bibinfo {year} {2010})}\BibitemShut {NoStop}%
\bibitem [{\citenamefont {Jiang}\ \emph {et~al.}(2009)\citenamefont {Jiang},
  \citenamefont {Lin}, \citenamefont {Rosenberg}, \citenamefont {Vahala},\ and\
  \citenamefont {Painter}}]{Jiang2009High}%
  \BibitemOpen
  \bibfield  {author} {\bibinfo {author} {\bibfnamefont {X.}~\bibnamefont
  {Jiang}}, \bibinfo {author} {\bibfnamefont {Q.}~\bibnamefont {Lin}}, \bibinfo
  {author} {\bibfnamefont {J.}~\bibnamefont {Rosenberg}}, \bibinfo {author}
  {\bibfnamefont {K.}~\bibnamefont {Vahala}}, \ and\ \bibinfo {author}
  {\bibfnamefont {O.}~\bibnamefont {Painter}},\ }\href {\doibase
  10.1364/OE.17.020911} {\bibfield  {journal} {\bibinfo  {journal} {Opt.
  Express}\ }\textbf {\bibinfo {volume} {17}},\ \bibinfo {pages} {20911}
  (\bibinfo {year} {2009})}\BibitemShut {NoStop}%
\bibitem [{\citenamefont {Wiederhecker}\ \emph {et~al.}(2009)\citenamefont
  {Wiederhecker}, \citenamefont {Chen}, \citenamefont {Gondarenko},\ and\
  \citenamefont {Lipson}}]{wiederhecker2009controlling}%
  \BibitemOpen
  \bibfield  {author} {\bibinfo {author} {\bibfnamefont {G.~S.}\ \bibnamefont
  {Wiederhecker}}, \bibinfo {author} {\bibfnamefont {L.}~\bibnamefont {Chen}},
  \bibinfo {author} {\bibfnamefont {A.}~\bibnamefont {Gondarenko}}, \ and\
  \bibinfo {author} {\bibfnamefont {M.}~\bibnamefont {Lipson}},\ }\href
  {http://dx.doi.org/10.1038/nature08584} {\bibfield  {journal} {\bibinfo
  {journal} {Nature (London)}\ }\textbf {\bibinfo {volume} {462}},\ \bibinfo
  {pages} {633} (\bibinfo {year} {2009})}\BibitemShut {NoStop}%
\bibitem [{\citenamefont {Ma}\ \emph {et~al.}(2007)\citenamefont {Ma},
  \citenamefont {Schliesser}, \citenamefont {Del'Haye}, \citenamefont
  {Dabirian}, \citenamefont {Anetsberger},\ and\ \citenamefont
  {Kippenberg}}]{ma2007radiation-pressure-driven}%
  \BibitemOpen
  \bibfield  {author} {\bibinfo {author} {\bibfnamefont {R.}~\bibnamefont
  {Ma}}, \bibinfo {author} {\bibfnamefont {A.}~\bibnamefont {Schliesser}},
  \bibinfo {author} {\bibfnamefont {P.}~\bibnamefont {Del'Haye}}, \bibinfo
  {author} {\bibfnamefont {A.}~\bibnamefont {Dabirian}}, \bibinfo {author}
  {\bibfnamefont {G.}~\bibnamefont {Anetsberger}}, \ and\ \bibinfo {author}
  {\bibfnamefont {T.~J.}\ \bibnamefont {Kippenberg}},\ }\href {\doibase
  10.1364/OL.32.002200} {\bibfield  {journal} {\bibinfo  {journal} {Opt.
  Lett.}\ }\textbf {\bibinfo {volume} {32}},\ \bibinfo {pages} {2200} (\bibinfo
  {year} {2007})}\BibitemShut {NoStop}%
\bibitem [{\citenamefont {Park}\ and\ \citenamefont
  {Wang}(2009)}]{park2009resolved-sideband}%
  \BibitemOpen
  \bibfield  {author} {\bibinfo {author} {\bibfnamefont {Y.~S.}\ \bibnamefont
  {Park}}\ and\ \bibinfo {author} {\bibfnamefont {H.~L.}\ \bibnamefont
  {Wang}},\ }\href {http://dx.doi.org/10.1038/nphys1303} {\bibfield  {journal}
  {\bibinfo  {journal} {Nat. Phys.}\ }\textbf {\bibinfo {volume} {5}},\
  \bibinfo {pages} {489} (\bibinfo {year} {2009})}\BibitemShut {NoStop}%
\bibitem [{\citenamefont {Thompson}\ \emph {et~al.}(2008)\citenamefont
  {Thompson}, \citenamefont {Zwickl}, \citenamefont {Jayich}, \citenamefont
  {Marquardt}, \citenamefont {Girvin},\ and\ \citenamefont
  {Harris}}]{thompson2007strong}%
  \BibitemOpen
  \bibfield  {author} {\bibinfo {author} {\bibfnamefont {J.~D.}\ \bibnamefont
  {Thompson}}, \bibinfo {author} {\bibfnamefont {B.~M.}\ \bibnamefont
  {Zwickl}}, \bibinfo {author} {\bibfnamefont {A.~M.}\ \bibnamefont {Jayich}},
  \bibinfo {author} {\bibfnamefont {F.}~\bibnamefont {Marquardt}}, \bibinfo
  {author} {\bibfnamefont {S.~M.}\ \bibnamefont {Girvin}}, \ and\ \bibinfo
  {author} {\bibfnamefont {J.~G.~E.}\ \bibnamefont {Harris}},\ }\href
  {http://dx.doi.org/10.1038/nature06715} {\bibfield  {journal} {\bibinfo
  {journal} {Nature (London)}\ }\textbf {\bibinfo {volume} {452}},\ \bibinfo
  {pages} {72} (\bibinfo {year} {2008})}\BibitemShut {NoStop}%
\bibitem [{\citenamefont {Favero}\ \emph {et~al.}(2009)\citenamefont {Favero},
  \citenamefont {Stapfner}, \citenamefont {Hunger}, \citenamefont
  {Paulitschke}, \citenamefont {Reichel}, \citenamefont {Lorenz}, \citenamefont
  {Weig},\ and\ \citenamefont {Karrai}}]{favero2009fluctuating}%
  \BibitemOpen
  \bibfield  {author} {\bibinfo {author} {\bibfnamefont {I.}~\bibnamefont
  {Favero}}, \bibinfo {author} {\bibfnamefont {S.}~\bibnamefont {Stapfner}},
  \bibinfo {author} {\bibfnamefont {D.}~\bibnamefont {Hunger}}, \bibinfo
  {author} {\bibfnamefont {P.}~\bibnamefont {Paulitschke}}, \bibinfo {author}
  {\bibfnamefont {J.}~\bibnamefont {Reichel}}, \bibinfo {author} {\bibfnamefont
  {H.}~\bibnamefont {Lorenz}}, \bibinfo {author} {\bibfnamefont {E.~M.}\
  \bibnamefont {Weig}}, \ and\ \bibinfo {author} {\bibfnamefont
  {K.}~\bibnamefont {Karrai}},\ }\href {\doibase 10.1364/OE.17.012813}
  {\bibfield  {journal} {\bibinfo  {journal} {Opt. Express}\ }\textbf {\bibinfo
  {volume} {17}},\ \bibinfo {pages} {12813} (\bibinfo {year}
  {2009})}\BibitemShut {NoStop}%
\bibitem [{\citenamefont {Regal}\ \emph {et~al.}(2008)\citenamefont {Regal},
  \citenamefont {Teufel},\ and\ \citenamefont {Lehnert}}]{regal2008measuring}%
  \BibitemOpen
  \bibfield  {author} {\bibinfo {author} {\bibfnamefont {C.~A.}\ \bibnamefont
  {Regal}}, \bibinfo {author} {\bibfnamefont {J.~D.}\ \bibnamefont {Teufel}}, \
  and\ \bibinfo {author} {\bibfnamefont {K.~W.}\ \bibnamefont {Lehnert}},\
  }\href {http://dx.doi.org/10.1038/nphys974} {\bibfield  {journal} {\bibinfo
  {journal} {Nat. Phys.}\ }\textbf {\bibinfo {volume} {4}},\ \bibinfo {pages}
  {555} (\bibinfo {year} {2008})}\BibitemShut {NoStop}%
\bibitem [{\citenamefont {Lee}\ \emph {et~al.}(2010)\citenamefont {Lee},
  \citenamefont {McRae}, \citenamefont {Harris}, \citenamefont {Knittel},\ and\
  \citenamefont {Bowen}}]{lee2009cooling}%
  \BibitemOpen
  \bibfield  {author} {\bibinfo {author} {\bibfnamefont {K.~H.}\ \bibnamefont
  {Lee}}, \bibinfo {author} {\bibfnamefont {T.~G.}\ \bibnamefont {McRae}},
  \bibinfo {author} {\bibfnamefont {G.~I.}\ \bibnamefont {Harris}}, \bibinfo
  {author} {\bibfnamefont {J.}~\bibnamefont {Knittel}}, \ and\ \bibinfo
  {author} {\bibfnamefont {W.~P.}\ \bibnamefont {Bowen}},\ }\href {\doibase
  10.1103/PhysRevLett.104.123604} {\bibfield  {journal} {\bibinfo  {journal}
  {Phys. Rev. Lett.}\ }\textbf {\bibinfo {volume} {104}},\ \bibinfo {pages}
  {123604} (\bibinfo {year} {2010})}\BibitemShut {NoStop}%
\bibitem [{\citenamefont {Winger}\ \emph {et~al.}(2011)\citenamefont {Winger},
  \citenamefont {Blasius}, \citenamefont {Alegre}, \citenamefont
  {Safavi-Naeini}, \citenamefont {Meenehan}, \citenamefont {Cohen},
  \citenamefont {Stobbe},\ and\ \citenamefont {Painter}}]{winger2011a}%
  \BibitemOpen
  \bibfield  {author} {\bibinfo {author} {\bibfnamefont {M.}~\bibnamefont
  {Winger}}, \bibinfo {author} {\bibfnamefont {T.~D.}\ \bibnamefont {Blasius}},
  \bibinfo {author} {\bibfnamefont {T.~P.~M.}\ \bibnamefont {Alegre}}, \bibinfo
  {author} {\bibfnamefont {A.~H.}\ \bibnamefont {Safavi-Naeini}}, \bibinfo
  {author} {\bibfnamefont {S.}~\bibnamefont {Meenehan}}, \bibinfo {author}
  {\bibfnamefont {J.}~\bibnamefont {Cohen}}, \bibinfo {author} {\bibfnamefont
  {S.}~\bibnamefont {Stobbe}}, \ and\ \bibinfo {author} {\bibfnamefont
  {O.}~\bibnamefont {Painter}},\ }\href {\doibase 10.1364/OE.19.024905}
  {\bibfield  {journal} {\bibinfo  {journal} {Opt. Express}\ }\textbf {\bibinfo
  {volume} {19}},\ \bibinfo {pages} {24905} (\bibinfo {year}
  {2011})}\BibitemShut {NoStop}%
\bibitem [{\citenamefont {Barzanjeh}\ \emph {et~al.}(2011)\citenamefont
  {Barzanjeh}, \citenamefont {Vitali}, \citenamefont {Tombesi},\ and\
  \citenamefont {Milburn}}]{barzanjeh2011entangling}%
  \BibitemOpen
  \bibfield  {author} {\bibinfo {author} {\bibfnamefont {S.}~\bibnamefont
  {Barzanjeh}}, \bibinfo {author} {\bibfnamefont {D.}~\bibnamefont {Vitali}},
  \bibinfo {author} {\bibfnamefont {P.}~\bibnamefont {Tombesi}}, \ and\
  \bibinfo {author} {\bibfnamefont {G.~J.}\ \bibnamefont {Milburn}},\ }\href
  {\doibase 10.1103/PhysRevA.84.042342} {\bibfield  {journal} {\bibinfo
  {journal} {Phys. Rev. A}\ }\textbf {\bibinfo {volume} {84}},\ \bibinfo
  {pages} {042342} (\bibinfo {year} {2011})}\BibitemShut {NoStop}%
\bibitem [{\citenamefont {Schmidt}\ \emph {et~al.}(2012)\citenamefont
  {Schmidt}, \citenamefont {Ludwig},\ and\ \citenamefont
  {Marquardt}}]{schmidt2012optomechanical}%
  \BibitemOpen
  \bibfield  {author} {\bibinfo {author} {\bibfnamefont {M.}~\bibnamefont
  {Schmidt}}, \bibinfo {author} {\bibfnamefont {M.}~\bibnamefont {Ludwig}}, \
  and\ \bibinfo {author} {\bibfnamefont {F.}~\bibnamefont {Marquardt}},\ }\href
  {http://stacks.iop.org/1367-2630/14/i=12/a=125005} {\bibfield  {journal}
  {\bibinfo  {journal} {New J. Phys.}\ }\textbf {\bibinfo {volume} {14}},\
  \bibinfo {pages} {125005} (\bibinfo {year} {2012})}\BibitemShut {NoStop}%
\bibitem [{\citenamefont {Mari}\ and\ \citenamefont
  {Eisert}(2009)}]{Mari2009Gently}%
  \BibitemOpen
  \bibfield  {author} {\bibinfo {author} {\bibfnamefont {A.}~\bibnamefont
  {Mari}}\ and\ \bibinfo {author} {\bibfnamefont {J.}~\bibnamefont {Eisert}},\
  }\href {\doibase 10.1103/PhysRevLett.103.213603} {\bibfield  {journal}
  {\bibinfo  {journal} {Phys. Rev. Lett.}\ }\textbf {\bibinfo {volume} {103}},\
  \bibinfo {pages} {213603} (\bibinfo {year} {2009})}\BibitemShut {NoStop}%
\bibitem [{\citenamefont {Mari}\ and\ \citenamefont
  {Eisert}(2012)}]{mari2012opto}%
  \BibitemOpen
  \bibfield  {author} {\bibinfo {author} {\bibfnamefont {A.}~\bibnamefont
  {Mari}}\ and\ \bibinfo {author} {\bibfnamefont {J.}~\bibnamefont {Eisert}},\
  }\href {http://stacks.iop.org/1367-2630/14/i=7/a=075014} {\bibfield
  {journal} {\bibinfo  {journal} {New J. Phys.}\ }\textbf {\bibinfo {volume}
  {14}},\ \bibinfo {pages} {075014} (\bibinfo {year} {2012})}\BibitemShut
  {NoStop}%
\bibitem [{\citenamefont {Abdi}\ and\ \citenamefont
  {Hartmann}(2015)}]{abdi2015entangling}%
  \BibitemOpen
  \bibfield  {author} {\bibinfo {author} {\bibfnamefont {M.}~\bibnamefont
  {Abdi}}\ and\ \bibinfo {author} {\bibfnamefont {M.~J.}\ \bibnamefont
  {Hartmann}},\ }\href {http://stacks.iop.org/1367-2630/17/i=1/a=013056}
  {\bibfield  {journal} {\bibinfo  {journal} {New J. Phys.}\ }\textbf {\bibinfo
  {volume} {17}},\ \bibinfo {pages} {013056} (\bibinfo {year}
  {2015})}\BibitemShut {NoStop}%
\bibitem [{\citenamefont {Li}\ \emph {et~al.}(2015{\natexlab{a}})\citenamefont
  {Li}, \citenamefont {Ma},\ and\ \citenamefont {Li}}]{li2015generations}%
  \BibitemOpen
  \bibfield  {author} {\bibinfo {author} {\bibfnamefont {Z.}~\bibnamefont
  {Li}}, \bibinfo {author} {\bibfnamefont {S.-l.}\ \bibnamefont {Ma}}, \ and\
  \bibinfo {author} {\bibfnamefont {F.-l.}\ \bibnamefont {Li}},\ }\href
  {\doibase 10.1103/PhysRevA.92.023856} {\bibfield  {journal} {\bibinfo
  {journal} {Phys. Rev. A}\ }\textbf {\bibinfo {volume} {92}},\ \bibinfo
  {pages} {023856} (\bibinfo {year} {2015}{\natexlab{a}})}\BibitemShut
  {NoStop}%
\bibitem [{\citenamefont {Wang}\ \emph {et~al.}(2016)\citenamefont {Wang},
  \citenamefont {L\"u}, \citenamefont {Wang}, \citenamefont {You},\ and\
  \citenamefont {Wu}}]{wang2016macroscopic}%
  \BibitemOpen
  \bibfield  {author} {\bibinfo {author} {\bibfnamefont {M.}~\bibnamefont
  {Wang}}, \bibinfo {author} {\bibfnamefont {X.-Y.}\ \bibnamefont {L\"u}},
  \bibinfo {author} {\bibfnamefont {Y.-D.}\ \bibnamefont {Wang}}, \bibinfo
  {author} {\bibfnamefont {J.~Q.}\ \bibnamefont {You}}, \ and\ \bibinfo
  {author} {\bibfnamefont {Y.}~\bibnamefont {Wu}},\ }\href {\doibase
  10.1103/PhysRevA.94.053807} {\bibfield  {journal} {\bibinfo  {journal} {Phys.
  Rev. A}\ }\textbf {\bibinfo {volume} {94}},\ \bibinfo {pages} {053807}
  (\bibinfo {year} {2016})}\BibitemShut {NoStop}%
\bibitem [{\citenamefont {Chen}\ \emph {et~al.}(2014)\citenamefont {Chen},
  \citenamefont {Shen}, \citenamefont {Yang}, \citenamefont {Wu},\ and\
  \citenamefont {Zheng}}]{chen2014enhancement}%
  \BibitemOpen
  \bibfield  {author} {\bibinfo {author} {\bibfnamefont {R.-X.}\ \bibnamefont
  {Chen}}, \bibinfo {author} {\bibfnamefont {L.-T.}\ \bibnamefont {Shen}},
  \bibinfo {author} {\bibfnamefont {Z.-B.}\ \bibnamefont {Yang}}, \bibinfo
  {author} {\bibfnamefont {H.-Z.}\ \bibnamefont {Wu}}, \ and\ \bibinfo {author}
  {\bibfnamefont {S.-B.}\ \bibnamefont {Zheng}},\ }\href {\doibase
  10.1103/PhysRevA.89.023843} {\bibfield  {journal} {\bibinfo  {journal} {Phys.
  Rev. A}\ }\textbf {\bibinfo {volume} {89}},\ \bibinfo {pages} {023843}
  (\bibinfo {year} {2014})}\BibitemShut {NoStop}%
\bibitem [{\citenamefont {Chen}\ \emph {et~al.}(2017)\citenamefont {Chen},
  \citenamefont {Liao},\ and\ \citenamefont {Lin}}]{sr2017}%
  \BibitemOpen
  \bibfield  {author} {\bibinfo {author} {\bibfnamefont {R.~X.}\ \bibnamefont
  {Chen}}, \bibinfo {author} {\bibfnamefont {C.~G.}\ \bibnamefont {Liao}}, \
  and\ \bibinfo {author} {\bibfnamefont {X.~M.}\ \bibnamefont {Lin}},\ }\href
  {https://doi.org/10.1038/s41598-017-15032-1} {\bibfield  {journal} {\bibinfo
  {journal} {Sci. Rep.}\ }\textbf {\bibinfo {volume} {7}},\ \bibinfo {pages}
  {14497} (\bibinfo {year} {2017})}\BibitemShut {NoStop}%
\bibitem [{\citenamefont {Wang}\ and\ \citenamefont
  {Clerk}(2013)}]{wang2013reservoir}%
  \BibitemOpen
  \bibfield  {author} {\bibinfo {author} {\bibfnamefont {Y.-D.}\ \bibnamefont
  {Wang}}\ and\ \bibinfo {author} {\bibfnamefont {A.~A.}\ \bibnamefont
  {Clerk}},\ }\href {\doibase 10.1103/PhysRevLett.110.253601} {\bibfield
  {journal} {\bibinfo  {journal} {Phys. Rev. Lett.}\ }\textbf {\bibinfo
  {volume} {110}},\ \bibinfo {pages} {253601} (\bibinfo {year}
  {2013})}\BibitemShut {NoStop}%
\bibitem [{\citenamefont {Wang}\ \emph {et~al.}(2015)\citenamefont {Wang},
  \citenamefont {Chesi},\ and\ \citenamefont {Clerk}}]{wang2015bipartite}%
  \BibitemOpen
  \bibfield  {author} {\bibinfo {author} {\bibfnamefont {Y.-D.}\ \bibnamefont
  {Wang}}, \bibinfo {author} {\bibfnamefont {S.}~\bibnamefont {Chesi}}, \ and\
  \bibinfo {author} {\bibfnamefont {A.~A.}\ \bibnamefont {Clerk}},\ }\href
  {\doibase 10.1103/PhysRevA.91.013807} {\bibfield  {journal} {\bibinfo
  {journal} {Phys. Rev. A}\ }\textbf {\bibinfo {volume} {91}},\ \bibinfo
  {pages} {013807} (\bibinfo {year} {2015})}\BibitemShut {NoStop}%
\bibitem [{\citenamefont {Tan}\ \emph {et~al.}(2013)\citenamefont {Tan},
  \citenamefont {Li},\ and\ \citenamefont {Meystre}}]{tan2013dissipation}%
  \BibitemOpen
  \bibfield  {author} {\bibinfo {author} {\bibfnamefont {H.}~\bibnamefont
  {Tan}}, \bibinfo {author} {\bibfnamefont {G.}~\bibnamefont {Li}}, \ and\
  \bibinfo {author} {\bibfnamefont {P.}~\bibnamefont {Meystre}},\ }\href
  {\doibase 10.1103/PhysRevA.87.033829} {\bibfield  {journal} {\bibinfo
  {journal} {Phys. Rev. A}\ }\textbf {\bibinfo {volume} {87}},\ \bibinfo
  {pages} {033829} (\bibinfo {year} {2013})}\BibitemShut {NoStop}%
\bibitem [{\citenamefont {Woolley}\ and\ \citenamefont
  {Clerk}(2014)}]{woolley2014two}%
  \BibitemOpen
  \bibfield  {author} {\bibinfo {author} {\bibfnamefont {M.~J.}\ \bibnamefont
  {Woolley}}\ and\ \bibinfo {author} {\bibfnamefont {A.~A.}\ \bibnamefont
  {Clerk}},\ }\href {\doibase 10.1103/PhysRevA.89.063805} {\bibfield  {journal}
  {\bibinfo  {journal} {Phys. Rev. A}\ }\textbf {\bibinfo {volume} {89}},\
  \bibinfo {pages} {063805} (\bibinfo {year} {2014})}\BibitemShut {NoStop}%
\bibitem [{\citenamefont {Yang}\ \emph {et~al.}(2015)\citenamefont {Yang},
  \citenamefont {An}, \citenamefont {Yang},\ and\ \citenamefont
  {Li}}]{yang2015generation}%
  \BibitemOpen
  \bibfield  {author} {\bibinfo {author} {\bibfnamefont {C.-J.}\ \bibnamefont
  {Yang}}, \bibinfo {author} {\bibfnamefont {J.-H.}\ \bibnamefont {An}},
  \bibinfo {author} {\bibfnamefont {W.}~\bibnamefont {Yang}}, \ and\ \bibinfo
  {author} {\bibfnamefont {Y.}~\bibnamefont {Li}},\ }\href {\doibase
  10.1103/PhysRevA.92.062311} {\bibfield  {journal} {\bibinfo  {journal} {Phys.
  Rev. A}\ }\textbf {\bibinfo {volume} {92}},\ \bibinfo {pages} {062311}
  (\bibinfo {year} {2015})}\BibitemShut {NoStop}%
\bibitem [{\citenamefont {Li}\ \emph {et~al.}(2015{\natexlab{b}})\citenamefont
  {Li}, \citenamefont {Haghighi}, \citenamefont {Malossi}, \citenamefont
  {Zippilli},\ and\ \citenamefont {Vitali}}]{Li2015Generation}%
  \BibitemOpen
  \bibfield  {author} {\bibinfo {author} {\bibfnamefont {J.}~\bibnamefont
  {Li}}, \bibinfo {author} {\bibfnamefont {I.~M.}\ \bibnamefont {Haghighi}},
  \bibinfo {author} {\bibfnamefont {N.}~\bibnamefont {Malossi}}, \bibinfo
  {author} {\bibfnamefont {S.}~\bibnamefont {Zippilli}}, \ and\ \bibinfo
  {author} {\bibfnamefont {D.}~\bibnamefont {Vitali}},\ }\href
  {http://stacks.iop.org/1367-2630/17/i=10/a=103037} {\bibfield  {journal}
  {\bibinfo  {journal} {New J. Phys.}\ }\textbf {\bibinfo {volume} {17}},\
  \bibinfo {pages} {103037} (\bibinfo {year} {2015}{\natexlab{b}})}\BibitemShut
  {NoStop}%
\bibitem [{\citenamefont {Qu}\ and\ \citenamefont {Agarwal}(2014)}]{Kenan}%
  \BibitemOpen
  \bibfield  {author} {\bibinfo {author} {\bibfnamefont {K.}~\bibnamefont
  {Qu}}\ and\ \bibinfo {author} {\bibfnamefont {G.~S.}\ \bibnamefont
  {Agarwal}},\ }\href {http://stacks.iop.org/1367-2630/16/i=11/a=113004}
  {\bibfield  {journal} {\bibinfo  {journal} {New Journal of Physics}\ }\textbf
  {\bibinfo {volume} {16}},\ \bibinfo {pages} {113004} (\bibinfo {year}
  {2014})}\BibitemShut {NoStop}%
\bibitem [{\citenamefont {Chen}\ \emph {et~al.}(2015)\citenamefont {Chen},
  \citenamefont {Shen},\ and\ \citenamefont {Zheng}}]{chen2015dissipation}%
  \BibitemOpen
  \bibfield  {author} {\bibinfo {author} {\bibfnamefont {R.-X.}\ \bibnamefont
  {Chen}}, \bibinfo {author} {\bibfnamefont {L.-T.}\ \bibnamefont {Shen}}, \
  and\ \bibinfo {author} {\bibfnamefont {S.-B.}\ \bibnamefont {Zheng}},\ }\href
  {\doibase 10.1103/PhysRevA.91.022326} {\bibfield  {journal} {\bibinfo
  {journal} {Phys. Rev. A}\ }\textbf {\bibinfo {volume} {91}},\ \bibinfo
  {pages} {022326} (\bibinfo {year} {2015})}\BibitemShut {NoStop}%
\bibitem [{\citenamefont {Ockeloen-Korppi}\ \emph {et~al.}(2017)\citenamefont
  {Ockeloen-Korppi}, \citenamefont {Damskagg}, \citenamefont {Pirkkalainen},
  \citenamefont {Clerk}, \citenamefont {Massel}, \citenamefont {Woolley},\ and\
  \citenamefont {Sillanpaa}}]{arxiv}%
  \BibitemOpen
  \bibfield  {author} {\bibinfo {author} {\bibfnamefont {C.~F.}\ \bibnamefont
  {Ockeloen-Korppi}}, \bibinfo {author} {\bibfnamefont {E.}~\bibnamefont
  {Damskagg}}, \bibinfo {author} {\bibfnamefont {J.~M.}\ \bibnamefont
  {Pirkkalainen}}, \bibinfo {author} {\bibfnamefont {A.~A.}\ \bibnamefont
  {Clerk}}, \bibinfo {author} {\bibfnamefont {F.}~\bibnamefont {Massel}},
  \bibinfo {author} {\bibfnamefont {M.~J.}\ \bibnamefont {Woolley}}, \ and\
  \bibinfo {author} {\bibfnamefont {M.~A.}\ \bibnamefont {Sillanpaa}},\ }\href
  {https://arxiv.org/abs/1711.01640} {\bibfield  {journal} {\bibinfo  {journal}
  {arXiv:1711.01640}\ } (\bibinfo {year} {2017})}\BibitemShut {NoStop}%
\bibitem [{\citenamefont {Vitali}\ \emph {et~al.}(2007)\citenamefont {Vitali},
  \citenamefont {Gigan}, \citenamefont {Ferreira}, \citenamefont {B\"ohm},
  \citenamefont {Tombesi}, \citenamefont {Guerreiro}, \citenamefont {Vedral},
  \citenamefont {Zeilinger},\ and\ \citenamefont
  {Aspelmeyer}}]{vitali2007optomechanical}%
  \BibitemOpen
  \bibfield  {author} {\bibinfo {author} {\bibfnamefont {D.}~\bibnamefont
  {Vitali}}, \bibinfo {author} {\bibfnamefont {S.}~\bibnamefont {Gigan}},
  \bibinfo {author} {\bibfnamefont {A.}~\bibnamefont {Ferreira}}, \bibinfo
  {author} {\bibfnamefont {H.~R.}\ \bibnamefont {B\"ohm}}, \bibinfo {author}
  {\bibfnamefont {P.}~\bibnamefont {Tombesi}}, \bibinfo {author} {\bibfnamefont
  {A.}~\bibnamefont {Guerreiro}}, \bibinfo {author} {\bibfnamefont
  {V.}~\bibnamefont {Vedral}}, \bibinfo {author} {\bibfnamefont
  {A.}~\bibnamefont {Zeilinger}}, \ and\ \bibinfo {author} {\bibfnamefont
  {M.}~\bibnamefont {Aspelmeyer}},\ }\href {\doibase
  10.1103/PhysRevLett.98.030405} {\bibfield  {journal} {\bibinfo  {journal}
  {Phys. Rev. Lett.}\ }\textbf {\bibinfo {volume} {98}},\ \bibinfo {pages}
  {030405} (\bibinfo {year} {2007})}\BibitemShut {NoStop}%
\bibitem [{\citenamefont {Paternostro}\ \emph {et~al.}(2007)\citenamefont
  {Paternostro}, \citenamefont {Vitali}, \citenamefont {Gigan}, \citenamefont
  {Kim}, \citenamefont {Brukner}, \citenamefont {Eisert},\ and\ \citenamefont
  {Aspelmeyer}}]{paternostro2007creating}%
  \BibitemOpen
  \bibfield  {author} {\bibinfo {author} {\bibfnamefont {M.}~\bibnamefont
  {Paternostro}}, \bibinfo {author} {\bibfnamefont {D.}~\bibnamefont {Vitali}},
  \bibinfo {author} {\bibfnamefont {S.}~\bibnamefont {Gigan}}, \bibinfo
  {author} {\bibfnamefont {M.~S.}\ \bibnamefont {Kim}}, \bibinfo {author}
  {\bibfnamefont {C.}~\bibnamefont {Brukner}}, \bibinfo {author} {\bibfnamefont
  {J.}~\bibnamefont {Eisert}}, \ and\ \bibinfo {author} {\bibfnamefont
  {M.}~\bibnamefont {Aspelmeyer}},\ }\href {\doibase
  10.1103/PhysRevLett.99.250401} {\bibfield  {journal} {\bibinfo  {journal}
  {Phys. Rev. Lett.}\ }\textbf {\bibinfo {volume} {99}},\ \bibinfo {pages}
  {250401} (\bibinfo {year} {2007})}\BibitemShut {NoStop}%
\bibitem [{\citenamefont {Braunstein}\ and\ \citenamefont
  {Kimble}(1998)}]{braunstein1998teleportation}%
  \BibitemOpen
  \bibfield  {author} {\bibinfo {author} {\bibfnamefont {S.~L.}\ \bibnamefont
  {Braunstein}}\ and\ \bibinfo {author} {\bibfnamefont {H.~J.}\ \bibnamefont
  {Kimble}},\ }\href {\doibase 10.1103/PhysRevLett.80.869} {\bibfield
  {journal} {\bibinfo  {journal} {Phys. Rev. Lett.}\ }\textbf {\bibinfo
  {volume} {80}},\ \bibinfo {pages} {869} (\bibinfo {year} {1998})}\BibitemShut
  {NoStop}%
\bibitem [{\citenamefont {Adesso}\ and\ \citenamefont
  {Illuminati}(2005)}]{adesso2005equivalence}%
  \BibitemOpen
  \bibfield  {author} {\bibinfo {author} {\bibfnamefont {G.}~\bibnamefont
  {Adesso}}\ and\ \bibinfo {author} {\bibfnamefont {F.}~\bibnamefont
  {Illuminati}},\ }\href {\doibase 10.1103/PhysRevLett.95.150503} {\bibfield
  {journal} {\bibinfo  {journal} {Phys. Rev. Lett.}\ }\textbf {\bibinfo
  {volume} {95}},\ \bibinfo {pages} {150503} (\bibinfo {year}
  {2005})}\BibitemShut {NoStop}%
\bibitem [{\citenamefont {Okamoto}\ \emph {et~al.}(2013)\citenamefont
  {Okamoto}, \citenamefont {Gourgout}, \citenamefont {Chang}, \citenamefont
  {Onomitsu}, \citenamefont {Mahboob}, \citenamefont {Chang},\ and\
  \citenamefont {Yamaguchi}}]{okamoto2013coherent}%
  \BibitemOpen
  \bibfield  {author} {\bibinfo {author} {\bibfnamefont {H.}~\bibnamefont
  {Okamoto}}, \bibinfo {author} {\bibfnamefont {A.}~\bibnamefont {Gourgout}},
  \bibinfo {author} {\bibfnamefont {C.-Y.}\ \bibnamefont {Chang}}, \bibinfo
  {author} {\bibfnamefont {K.}~\bibnamefont {Onomitsu}}, \bibinfo {author}
  {\bibfnamefont {I.}~\bibnamefont {Mahboob}}, \bibinfo {author} {\bibfnamefont
  {E.~Y.}\ \bibnamefont {Chang}}, \ and\ \bibinfo {author} {\bibfnamefont
  {H.}~\bibnamefont {Yamaguchi}},\ }\href {http://dx.doi.org/10.1038/nphys2665}
  {\bibfield  {journal} {\bibinfo  {journal} {Nat. Phys.}\ }\textbf {\bibinfo
  {volume} {9}},\ \bibinfo {pages} {480} (\bibinfo {year} {2013})}\BibitemShut
  {NoStop}%
\bibitem [{\citenamefont {Buks}\ and\ \citenamefont
  {Roukes}(2002)}]{buks2002electrically}%
  \BibitemOpen
  \bibfield  {author} {\bibinfo {author} {\bibfnamefont {E.}~\bibnamefont
  {Buks}}\ and\ \bibinfo {author} {\bibfnamefont {M.~L.}\ \bibnamefont
  {Roukes}},\ }\href@noop {} {\bibfield  {journal} {\bibinfo  {journal} {J.
  Microelectromech. Syst.}\ }\textbf {\bibinfo {volume} {11}},\ \bibinfo
  {pages} {802} (\bibinfo {year} {2002})}\BibitemShut {NoStop}%
\bibitem [{\citenamefont {Hensinger}\ \emph {et~al.}(2005)\citenamefont
  {Hensinger}, \citenamefont {Utami}, \citenamefont {Goan}, \citenamefont
  {Schwab}, \citenamefont {Monroe},\ and\ \citenamefont
  {Milburn}}]{Hensinger2005Ion}%
  \BibitemOpen
  \bibfield  {author} {\bibinfo {author} {\bibfnamefont {W.~K.}\ \bibnamefont
  {Hensinger}}, \bibinfo {author} {\bibfnamefont {D.~W.}\ \bibnamefont
  {Utami}}, \bibinfo {author} {\bibfnamefont {H.-S.}\ \bibnamefont {Goan}},
  \bibinfo {author} {\bibfnamefont {K.}~\bibnamefont {Schwab}}, \bibinfo
  {author} {\bibfnamefont {C.}~\bibnamefont {Monroe}}, \ and\ \bibinfo {author}
  {\bibfnamefont {G.~J.}\ \bibnamefont {Milburn}},\ }\href {\doibase
  10.1103/PhysRevA.72.041405} {\bibfield  {journal} {\bibinfo  {journal} {Phys.
  Rev. A}\ }\textbf {\bibinfo {volume} {72}},\ \bibinfo {pages} {041405}
  (\bibinfo {year} {2005})}\BibitemShut {NoStop}%
\bibitem [{\citenamefont {Zhang}\ \emph {et~al.}(2012)\citenamefont {Zhang},
  \citenamefont {Li}, \citenamefont {Feng},\ and\ \citenamefont
  {Xu}}]{Zhang2012Precision}%
  \BibitemOpen
  \bibfield  {author} {\bibinfo {author} {\bibfnamefont {J.-Q.}\ \bibnamefont
  {Zhang}}, \bibinfo {author} {\bibfnamefont {Y.}~\bibnamefont {Li}}, \bibinfo
  {author} {\bibfnamefont {M.}~\bibnamefont {Feng}}, \ and\ \bibinfo {author}
  {\bibfnamefont {Y.}~\bibnamefont {Xu}},\ }\href {\doibase
  10.1103/PhysRevA.86.053806} {\bibfield  {journal} {\bibinfo  {journal} {Phys.
  Rev. A}\ }\textbf {\bibinfo {volume} {86}},\ \bibinfo {pages} {053806}
  (\bibinfo {year} {2012})}\BibitemShut {NoStop}%
\bibitem [{\citenamefont {Ma}\ \emph {et~al.}(2014)\citenamefont {Ma},
  \citenamefont {Zhang}, \citenamefont {Xiao}, \citenamefont {Feng},\ and\
  \citenamefont {Zhang}}]{Ma2014Tunable}%
  \BibitemOpen
  \bibfield  {author} {\bibinfo {author} {\bibfnamefont {P.-C.}\ \bibnamefont
  {Ma}}, \bibinfo {author} {\bibfnamefont {J.-Q.}\ \bibnamefont {Zhang}},
  \bibinfo {author} {\bibfnamefont {Y.}~\bibnamefont {Xiao}}, \bibinfo {author}
  {\bibfnamefont {M.}~\bibnamefont {Feng}}, \ and\ \bibinfo {author}
  {\bibfnamefont {Z.-M.}\ \bibnamefont {Zhang}},\ }\href {\doibase
  10.1103/PhysRevA.90.043825} {\bibfield  {journal} {\bibinfo  {journal} {Phys.
  Rev. A}\ }\textbf {\bibinfo {volume} {90}},\ \bibinfo {pages} {043825}
  (\bibinfo {year} {2014})}\BibitemShut {NoStop}%
\bibitem [{\citenamefont {Adesso}\ and\ \citenamefont
  {Illuminati}(2007)}]{adesso2007entanglement}%
  \BibitemOpen
  \bibfield  {author} {\bibinfo {author} {\bibfnamefont {G.}~\bibnamefont
  {Adesso}}\ and\ \bibinfo {author} {\bibfnamefont {F.}~\bibnamefont
  {Illuminati}},\ }\href {http://stacks.iop.org/1751-8121/40/i=28/a=S01}
  {\bibfield  {journal} {\bibinfo  {journal} {J. Phys. A}\ }\textbf {\bibinfo
  {volume} {40}},\ \bibinfo {pages} {7821} (\bibinfo {year}
  {2007})}\BibitemShut {NoStop}%
\bibitem [{\citenamefont {Weedbrook}\ \emph {et~al.}(2012)\citenamefont
  {Weedbrook}, \citenamefont {Pirandola}, \citenamefont {Garc\'{\i}a-Patr\'on},
  \citenamefont {Cerf}, \citenamefont {Ralph}, \citenamefont {Shapiro},\ and\
  \citenamefont {Lloyd}}]{weedbrook2012gaussian}%
  \BibitemOpen
  \bibfield  {author} {\bibinfo {author} {\bibfnamefont {C.}~\bibnamefont
  {Weedbrook}}, \bibinfo {author} {\bibfnamefont {S.}~\bibnamefont
  {Pirandola}}, \bibinfo {author} {\bibfnamefont {R.}~\bibnamefont
  {Garc\'{\i}a-Patr\'on}}, \bibinfo {author} {\bibfnamefont {N.~J.}\
  \bibnamefont {Cerf}}, \bibinfo {author} {\bibfnamefont {T.~C.}\ \bibnamefont
  {Ralph}}, \bibinfo {author} {\bibfnamefont {J.~H.}\ \bibnamefont {Shapiro}},
  \ and\ \bibinfo {author} {\bibfnamefont {S.}~\bibnamefont {Lloyd}},\ }\href
  {\doibase 10.1103/RevModPhys.84.621} {\bibfield  {journal} {\bibinfo
  {journal} {Rev. Mod. Phys.}\ }\textbf {\bibinfo {volume} {84}},\ \bibinfo
  {pages} {621} (\bibinfo {year} {2012})}\BibitemShut {NoStop}%
\bibitem [{\citenamefont {Olivares}\ and\ \citenamefont
  {Paris}(2012)}]{olivares2012quantum}%
  \BibitemOpen
  \bibfield  {author} {\bibinfo {author} {\bibfnamefont {S.}~\bibnamefont
  {Olivares}}\ and\ \bibinfo {author} {\bibfnamefont {M.~G.~A.}\ \bibnamefont
  {Paris}},\ }\href {https://doi.org/10.1140/epjst/e2012-01542-2} {\bibfield
  {journal} {\bibinfo  {journal} {Eur. Phys. J. Special Topics}\ }\textbf
  {\bibinfo {volume} {203}},\ \bibinfo {pages} {185} (\bibinfo {year}
  {2012})}\BibitemShut {NoStop}%
\bibitem [{\citenamefont {Teschl}(2012)}]{Teschl2012}%
  \BibitemOpen
  \bibfield  {author} {\bibinfo {author} {\bibfnamefont {G.}~\bibnamefont
  {Teschl}},\ }\href@noop {} {\emph {\bibinfo {title} {Ordinary differential
  equations and dynamical systems}}}\ (\bibinfo  {publisher} {Amer. Math.
  Soc.},\ \bibinfo {address} {Providence},\ \bibinfo {year} {2012})\BibitemShut
  {NoStop}%
\bibitem [{\citenamefont {Plenio}(2005)}]{plenio2005logarithmic}%
  \BibitemOpen
  \bibfield  {author} {\bibinfo {author} {\bibfnamefont {M.~B.}\ \bibnamefont
  {Plenio}},\ }\href {\doibase 10.1103/PhysRevLett.95.090503} {\bibfield
  {journal} {\bibinfo  {journal} {Phys. Rev. Lett.}\ }\textbf {\bibinfo
  {volume} {95}},\ \bibinfo {pages} {090503} (\bibinfo {year}
  {2005})}\BibitemShut {NoStop}%
\bibitem [{\citenamefont {Vidal}\ and\ \citenamefont
  {Werner}(2002)}]{vidal2002computable}%
  \BibitemOpen
  \bibfield  {author} {\bibinfo {author} {\bibfnamefont {G.}~\bibnamefont
  {Vidal}}\ and\ \bibinfo {author} {\bibfnamefont {R.~F.}\ \bibnamefont
  {Werner}},\ }\href {\doibase 10.1103/PhysRevA.65.032314} {\bibfield
  {journal} {\bibinfo  {journal} {Phys. Rev. A}\ }\textbf {\bibinfo {volume}
  {65}},\ \bibinfo {pages} {032314} (\bibinfo {year} {2002})}\BibitemShut
  {NoStop}%
\end{thebibliography}%
\bibliographystyle{apsrev4-1}

\end{document}